\documentclass[conference]{IEEEtran}
\IEEEoverridecommandlockouts
\usepackage{lipsum}% For this example
\usepackage{algorithm}
\usepackage{algpseudocode}
\usepackage{amsmath}
\usepackage{setspace}
\usepackage{graphicx} 
\usepackage{float} 
\usepackage{subfigure} 
\usepackage{multirow}
\usepackage{diagbox}
\usepackage{ulem}
\usepackage{color}
\usepackage{enumitem}
\usepackage{verbatim}
\usepackage{cleveref}
\usepackage{booktabs}
\usepackage{makecell}
\usepackage{amssymb}
\usepackage{caption}

\makeatletter
\newcounter{subfigure@save}
\makeatother

\usepackage{subfig}

\Crefformat{equation}{Equation~#2#1#3}

\def\BibTeX{{\rm B\kern-.05em{\sc i\kern-.025em b}\kern-.08em
    T\kern-.1667em\lower.7ex\hbox{E}\kern-.125emX}}
\begin{document}

\vspace{-3mm}
\title{Learning $k$-Determinantal Point Processes for \\ Personalized Ranking
%{\footnotesize \textsuperscript{*}Note: Sub-titles are not captured in Xplore andshould not be used}
\vspace{-1mm}
}
\author{\IEEEauthorblockN{Yuli Liu}
\IEEEauthorblockA{Qinghai University, China \& \\
Australian National University, Australia \\
liuyuli012@gmail.com
\vspace{-4mm}
}
\and
\IEEEauthorblockN{Christian Walder}
\IEEEauthorblockA{Google Brain \\
Montreal, Canada \\
cwalder@google.com
\vspace{-4mm}
}
\and
\IEEEauthorblockN{Lexing Xie}
\IEEEauthorblockA{Australian National University, Australia \& \\
Data61 CSIRO, Australia \\
lexing.xie@anu.edu.au
\vspace{-4mm}
}}
\maketitle

\begin{abstract}
The key to personalized recommendation is to predict a personalized ranking on a catalog of items by modeling the user’s preferences. 
There are many personalized ranking approaches for item recommendation from implicit feedback like Bayesian Personalized Ranking (BPR) and listwise ranking. 
Despite these methods have shown performance benefits, there are still limitations affecting recommendation performance. First, none of them directly optimize ranking of sets, causing inadequate exploitation of correlations among multiple items. 
Second, the diversity aspect of recommendations is insufficiently addressed compared to relevance. 

In this work, we present a new optimization criterion L$k$P based on set probability comparison for personalized ranking that moves beyond traditional ranking-based methods.
It formalizes set-level relevance and diversity ranking comparisons through a Determinantal Point Process (DPP) kernel decomposition. To confer ranking interpretability to the DPP set probabilities and prioritize the practicality of L$k$P, we condition the standard DPP on the cardinality $k$ of the DPP-distributed set, known as $k$-DPP, a less-explored extension of DPP. The generic stochastic gradient descent based technique can be directly applied to optimizing models that employ L$k$P. 
We implement L$k$P in the context of both Matrix Factorization (MF) and neural networks approaches, on three real-world datasets, obtaining improved relevance and diversity performances. 
L$k$P is broadly applicable, and when applied to existing recommendation models it also yields strong performance improvements, suggesting that L$k$P holds significant value to the field of recommender systems.
\end{abstract}

\vspace{-1mm}

\vspace{-1mm}
\section{Introduction}
\vspace{-1.2mm}
Recommender systems are ubiquitous services in many online platforms including social media and e-commerce, and play a crucial role in alleviating information overload by efficiently anticipating user preferences. 
%Personalization is attractive both for content providers, who can increase sales or views, and for customers, who can find interesting content more easily.
This work focuses on item recommendation \cite{rendle2012bpr, he2017neural}, wherein the task involves generating a personalized ranking of items based on the user's observed interactions with the system. 
In real-world scenarios, implicit feedback, such as clicks and views, is more readily available and collected compared to explicit feedback, \textit{e.g.}, ratings and reviews, making it a valuable resource for understanding user preferences.

To guide recommendation models in learning user preferences from implicit feedback, optimization objectives play a crucial role. 
One commonly used approach involves pointwise loss functions, \textit{e.g.}, binary cross entropy and squared error, which measure the discrepancy between the model's predicted preference scores and the actual labels on an individual instance basis.
%%Expanding on this, 
Given that the goal of item recommendation is to provide a personalized ranking list to a specific user, it naturally leads to the adoption of ranking-based optimization methods. Among these, 
%Bayesian Personalized Ranking (BPR), listwise models, and the newer setwise models are some of the principal techniques used in this domain.
%Considering the goal of item recommendation, \textit{i.e.}, providing a personalized ranking list to a specific user, ranking-based optimization , among which Bayesian Personalized Ranking (BPR), listwise, and recent setwise models \cite{chen2021set2setrank, wang2020setrank} are the most notable. 
%Pointwise learning is usually conducted by regression, to minimize the discrepancy between the predicted relevance $\hat{y}_{ui}$ and its target value ($1$ or $0$).%, \textit{i.e.}, via a squared loss, or by handling recommendation tasks as a binary classification problem and thereby strenthening the probability of the target item being recommended, namely, binary cross-entropy loss. 
the popular pairwise ranking approach, BPR \cite{rendle2012bpr}, assumes that an observed item ranks higher than a randomly selected unobserved (negative) item from the perspective of a specific user. 
Listwise ranking methods usually maximize the probability of permutation of the target list \cite{cao2007learning, xia2008listwise}. 
%However, two challenges stand out: there is no clear modeling of sequential relationship, rather only binary ratings are captured in the implicit feedback; typically, measuring the agreement between the Top-$N$ items on the observed and predicted lists results in exponential computational complexity in $N$ (a key but undesirable reason that $N$ is often set to 1). 
In recent years, some studies have attempted to bring the correlation within items into ranking-based optimization, leading to the topic of setwise ranking. Two eminent and state-of-the-art set-based ranking methods are SetRank \cite{wang2020setrank} and Set2SetRank \cite{chen2021set2setrank}. SetRank encourages an observed item to rank in front of multiple unobserved items in each list by making use of the concept of permutation probability. Set2SetRank models the set ranking based on item to item comparison and summarization of distance measurements between items. %%(a  of comparisons of individual items within the sets).
%(\romannumeral1) item to item comparison, \textit{i.e.}, encouraging each item of an observed set to rank higher than any items of the unobserved set, and (\romannumeral2) set distance measurement, \textit{i.e.}, reinforcing a margin of distance (summarized based on comparisons of items in sets) between observed and unobserved sets. 

%%The above-mentioned approaches indeed bring positive effects to recommendation models. 
%%However, we argue that two aspects that are important to recommender systems have not been adequately explored.
However, there are remain two problems in existing approaches: (\textit{\romannumeral1}) Structure information and correlations are far from well exploited and modeled. 
As shown in the upper layer of \Cref{fig-k-dpp:framework}, the correlations among items are fully ignored by pointwise and pairwise learning; listwise-based models center solely on the order correlation in the ranking list of items, while the binary nature of implicit feedback obstructs the acquisition of clear order correlations; the recently proposed setwise ranking optimization models (\textit{e.g.}, SetRank and Set2SetRank) intend to construct set-level comparisons, but are still confined to a BPR optimization criterion, and the set ranking is obtained by summarizing the individual items in each set, rather than recognizing multiple items as a whole (\textit{i.e.} set) that they are in order to more meaningfully compare them. Taking implicit feedback in \Cref{fig-k-dpp:framework} as an illustration, a ranking optimization, which treats entire sets as cohesive entities for comparison (concrete set-level optimization), not only enables us to model explicit ranking comparisons, such as {\fontsize{7}{7} $\left\{v_3^{(1)}, v_6^{(1)}, v_8^{(1)}\right\}>_u\left\{v_4^{(0)}, v_5^{(0)}, v_9^{(0)}\right\}$} (superscripts $1$ and $0$ denote observed and unobserved items, respectively), but also to discern subtle preferences, such as the greater potential interest of item {\fontsize{7}{7}$v_4^{(0)}$} over {\fontsize{7}{7}$v_5^{(0)}$} in {\fontsize{7.5}{7}$\left\{v_3^{(1)}, v_6^{(1)}, v_4^{(0)}\right\}>_u\left\{v_3^{(1)}, v_6^{(1)}, v_5^{(0)}\right\}$} and the stronger dependency of {\fontsize{7}{7}$v_8^1$} with {\fontsize{7}{7}$v_3^{(1)}$} and {\fontsize{7}{7}$v_6^{(1)}$} compared to {\fontsize{7}{7}$v_{10}^{(1)}$} in {\fontsize{7}{7}$\left\{v_3^{(1)}, v_6^{(1)}, v_8^{(1)}\right\}>_u\left\{v_3^{(1)}, v_6^{(1)}, v_{10}^{(1)}\right\}$}. This implies that the concrete set-level optimization criterion offers a wide range of potential combinations and aids in capturing intricate dependency relationships;
(\textit{\romannumeral2}) The objective of diversifying recommendations to mine additional interests for users, is not involved in existing optimization criterion approaches. Consequently, item recommendation models that employ these optimization objectives tend to favor popular and rather relevant items (often already known by users) \cite{wu2019pd, boim2011direc}, but lose sight of raising the ranking of diversified results for broadening users’ horizons.

To overcome these limitations, this work introduces a concrete set-level optimization criterion that facilitates ranking optimization through a direct comparison of different sets' probabilities of being a $k$-DPP. Here, $k$-DPP refers to a less-explored extension of the well-established concept, Determinantal Point Process (DPP). By leveraging the balance-focused quality \textit{vs.} diversity decomposition of the DPP kernel, we are able to seamlessly integrate two types of ranking comparison (\textit{i.e.}, \textit{relevance comparison} and \textit{diversity comparison}), thus offering a comprehensive measure of sets. The motivation for employing $k$-DPP lies in its ability to add ranking interpretability to DPP set probabilities, making them practically useful for achieving optimal recommendation sets. 
We name the proposed optimization criterion as L$k$P, an acronym underlining its aim to \textbf{l}earn \textbf{p}robabilities of specific subsets under the tailored $k$-DPP distribution. 
L$k$P presents an effective solution that acknowledges item correlations and diversity, thereby delivering an advanced recommendation model that boosts accuracy and diversity. 

We summarize the contributions of this work as follows: 
\begin{itemize}
%\item We provide a new perspective, \textit{i.e.} full-fledged set-level optimization criterion based on $k$-DPP set probability ranking, moving beyond the commonly used BPR metric, for item recommendation.
%\item The balanced $k$-DPP kernel enables recommendation models to learn diversity-aware embeddings for personalized ranking.  
\item We propose a concrete set-level optimization criterion, named L$k$P, that advances ranking optimization by allowing a direct comparison of set probabilities under a tailored $k$-DPP distribution, moving beyond the common BPR metric. Unlike prior approaches that focus on individual items, our method provides a comprehensive comparison at the set level.
\item We utilize $k$-DPP to provide ranking interpretability to DPP set probabilities, rendering them practically advantageous for ranking optimization. As a result, L$k$P criterion recognizes item correlations and diversity, thus bolstering the accuracy and diversity of recommendations.
\item By incorporating both the actual inclusion probability of an observed set and the exclusion probability of an unobserved set, we develop two distinct optimization approaches, namely L$k$P$_{PS}$ and L$k$P$_{NPS}$, which we introduce and compare.
%% \item The derivation of the gradient of L$k$P with respect to the model parameters demonstrates the feasibility of optimizing our approaches via stochastic gradient descent-based methods, thus enabling the maximization of L$k$P. 
%This strategy elegantly accommodates complex real-world scenarios and allows for practical, scalable implementation. 
\end{itemize}

%\begin{figure}
%  \centering
%  \includegraphics[width=.97\linewidth]{kdpp_sketch.png}
%  \caption{The diagrammatic sketch of L$k$P. Different colors mean different categories. \cwnote{Can you clarify this figure? It is surprising that the (1,8,9) and (1,3,5) indices etc change int he bottom to (1,4,6) etc. Where does (1,4,6) come from ? What is going on? What are you trying to show?}}
%  \Description{.}
%  \vspace{-4mm}
%\end{figure}
\vspace{-1mm}
\section{RELATED WORK}
\vspace{-1mm}
\label{sec-k-dpp:related-work}
%In this section, we review three lines of research that are highly relevant to this work. 
\textbf{Item Recommendation Model.}
As one of the most promising technologies to build recommender systems, collaborative filtering (CF) uses the known collaborative perception of users to make recommendations \cite{su2009survey, wu2016personal}. There are generally four varieties of CF models: matrix factorization (MF) learns user and item embeddings based on a low-rank decomposition of the user-item interaction matrix \cite{mnih2007probabilistic, koren2009matrix, koren2008factorization}; neural collaborative filtering replaces the inner product (a common notion of interaction in MF) with nonlinear neural networks \cite{he2017neural, li2020hierarchical}; collaborative deep learning couples deep representation learning for content information and collaborative filtering for feedback matrix \cite{wang2015collaborative, chen2020revisiting};  and translation-based CF adopts latent user-item relation vectors to model interactions \cite{tay2018latent, he2017translation}.
Among these models, Graph Convolutional Network (GCN), a category of collaborative deep learning, has gained dominance due to their effective modeling of item-item relationships through graph neural networks \cite{chen2022attentive, zhao2021variational}. 
%%Our proposed optimization criterion, L$k$P, is distinct from CF models that contribute to the architectural design of recommender systems. It serves as a guiding principle for optimizing the personalized ranking of recommendation sets. 
%L$k$P introduces a novel approach by directly comparing set probabilities using the k-DPP framework, which enables the exploration of both relevance and diversity aspects in item recommendations. 
%%By integrating the L$k$P based criterion into existing CF models like MF and GCN based CF methods, we enhance their ability to deliver more accurate and diverse recommendations.

\textbf{Ranking-based Optimization Criterion.}
BPR is the most popular pairwise approach for implicit feedback-based recommendation \cite{wan2022cross, gao2019neural}, which treats each item pair independently, and ignores the correlation of multiple observed items and multiple unobserved items. The listwise approaches measure the ranking loss by calculating the distance between predicted list and true list \cite{xia2008listwise, cao2007learning, shi2010list}.
SetRank \cite{wang2020setrank} provides a new research perspective for listwise learning with implicit feedback by using the permutation probability to encourage one observed item to rank in front of a set of unobserved items in each list. As this work focuses on ranking-based optimization with implicit feedback and SetRank has achieved state-of-the-art ranking results \cite{chen2021set2setrank, yu2020collaborative}, other list-based ranking-based methods are not compared. 
Although Set2SetRank \cite{chen2021set2setrank} aims to construct set-level ranking pairs by comparing the individual items in the set, it still uses the BPR optimization criterion.      
%The training efficiency of applying DPP-based likelihood has been evaluated, whose training time is near twice that of the BPR loss \cite{liu2022determinantal} . This is also verified in this work. 
%%In experiments, the same size of training instances as BCE and BPR is used.
%%Due to the ranking interpretation in $k$-DPP probability, comparing rankings of all subsets in a ground set is not required, which improves the training efficiency. In our experiments, we observed that on three separate datasets within a neural context, the time taken to achieve the best performance using L$k$P is approximately double compared to when using BPR loss.
%%Taking into account the substantial improvements in both quality and diversity brought about by $k$-DPP, this degree of training efficiency can be justified. 
%\vspace{1mm}

\textbf{DPP for Recommendation.} 
%In view of the ability of measuring set diversity, 
DPP has been a preferred choice of researchers to diversify recommendations. To relieve the computation complexity issue of maximum a posteriori (MAP) for DPP generation, Chen et al. \cite{chen2018fast} develop a novel algorithm to accelerate the MAP inference for DPP, which enables multiple studies \cite{wu2019pd, liu2020diversified, gan2020enhancing, gartrell2017low} to promote diversity by generating diversified recommendation lists. As these models balance diversity and relevance by blending DPP MAP inference into complex recommendation frameworks (\textit{e.g.}, GAN \cite{goodfellow2020generative} or knowledge graphs \cite{gan2020enhancing}) instead of focusing on ranking optimization approaches, we omit them in the Comparisons section. 
DPP is also used for basket completion tasks by viewing the basket completion task as a multi-class classification problem \cite{warlop2019tensorized}, in which the probability normalization and cardinality $k$ are ignored, making it inapplicable to traditional recommendation. 
%%The recent work \cite{liu2022determinantal} applies conditional DPP likelihoods in sequential recommendation to capture the dependency between previous items and targets. 
%%However, in line with the discussion in \Cref{sec-k-dpp:k-dpp-probability}, using standard DPP for item recommendation (a ranking optimization problem) makes the ranking interpretation unreasonable. 
%%According to our experiments under the same setting as \Cref{tab-k-dpp:GCN}-\Cref{tab-k-dpp:seminal}, applying standard DPP for ranking probability formulation achieves an unacceptable performance that is much weaker than BPR. This is the reason why the results based on standard DPP are not presented in Experiments.
\vspace{-1mm}
\section{Methodology}
\vspace{-1mm}
%%This work focuses on implicit feedback, such as clicking an item, purchasing a product, or checking-in to a location, as most user feedback is implicit in many real-world applications. Compared to explicit feedback (\textit{e.g.}, ratings and reviews), implicit feedback is more readily available, but it is more challenging to utilize, due to the natural scarcity of negative feedback \cite{he2017neural}.
%%In this section, the preliminary and L$k$P optimization criterion based on $k$-DPP set probability comparison for personalized ranking are successively presented.
In this section, we primarily focus on the L$k$P optimization criterion, preceded by related preliminaries.

  %%change this title
\vspace{-1mm}
\subsection{Preliminaries}  
\vspace{-1mm}
\subsubsection{Formulation} 
An item recommendation problem is generally defined given user set $\mathbb{U}$ and item set $\mathbb{V}$ with $N$ and $M$ as the size of $\mathbb{U}$ and $\mathbb{V}$, respectively.
Let $\mathbf{Y} \in \{0,1\}^{N\times M}$ denotes the user-item interaction matrix. 
%%Each element $y_{u v}$ in $\mathbf{Y}$ represents implicit feedback from user $u$ on item $v$. $y_{u v}=1$ and $y_{u v}=0$ indicate observed feedback (user $u$ likes item $v$) and unobserved feedback (user $u$ is not interested in item $v$), respectively. 
$Y_u^{+}$ indicates observed interactions of user $u$ and $Y_u^{-}=\mathbb{V} \backslash Y_u^{+}$ is the non-observed (negative) feedback. 
Given the interaction matrix $\mathbf{Y}$,
the goal of item recommendation is to anticipate each user $u$'s preference to item $v$ as $\hat{y}_{u v}$, typically with ranking-based optimization functions that are suitable for selecting items of likely interest to users.

In our vision of a concrete set-level optimization criterion L$k$P, we emphasize two ranking comparisons: (\textit{\romannumeral1}) The \textit{relevance comparison} $S_u >_u S_u^{\prime}$ (personalized for user $u$), where $S_u \subseteq Y_u^{+}$, $S_u^{\prime} \cap Y_u^{-} \neq \emptyset$, and $\left|S_u\right|=\left|S_u^{\prime}\right|=k$, represents that the user prefers a set filled with observed items over all other sets that contain at least one unobserved items; 
(\textit{\romannumeral2}) the \textit{diversity comparison} $|\mathcal{C}(S_u)|>_d |\mathcal{C}(S_u^{\prime})|$ (bears no personalization), where $\mathcal{C}(S_u)$ is the set of unique categories spanned by the items in $S_u$, means that a set is ranked higher if it spans more unique item categories/topics. Together, they aim to provide recommendations that balance personal relevance with diversity in category coverage. As this work focuses on optimization criterion design, we assume the recommendation model is available, such as matrix factorization models \cite{rendle2012bpr} or neural graph models \cite{wang2019neural}. To implement L$k$P, the input (training instance) comprised of a user ID and a $k$-DPP ground set with IDs of $k$ positive and $n$ negative items (the choices of $k$ and $n$ will be detailed in \Cref{sec-k-dpp:k-dpp-probability}) is provided at first. Once embedded, the fetched embeddings are then modeled and learned through an existing recommendation framework. The final output consists of predicted relevance scores (dot product of user-item embeddings \cite{liang2021recommending, rendle2012bpr} or binary classification result from neural networks \cite{he2017neural}) for $k+n$ items.
In L$k$P optimization, we integrate these scores (quality) with the category-related diversity measurement, culminating in a tailored $k$-DPP over entire $k$-sized subsets (\textit{i.e.}, $C(k+n,k)$ different combinations) of the $k+n$ ground set. 
%%Our objective is to maximize the target probability as dictated by the k-DPP. Given that a k-DPP distribution encompasses all k-sized subsets within the ground set, optimizing for a specific subset endows it with a k-DPP probability ranking interpretation.

\subsubsection{DPP} As an elegant probabilistic model capable of modeling repulsive correlations \cite{kulesza2012determinantal}, DPP has been widely studied in machine learning fields.
%quantum physics and random matrix theory for several decades, and more recently  in machine learning. 
%The strength of repulsion is parameterised by a kernel matrix that defines a global measure of similarity between pairs of items. 
Given a discrete set $\mathcal{S}=\{1,2, \ldots, M\}$ (\textit{e.g.}, item set $\mathbb{V}$), a DPP $P$ is a probability measure on $2^\mathcal{S}$, the set of all subsets of $\mathcal{S}$. When $P$ gives nonzero probability to empty set, there exists a matrix $\mathbf{L} \in \mathbb{R}^{M \times M}$ determining the probability of any subset $S$ of $\mathcal{S}$ as
\vspace{-0.5mm}
\begin{equation}
P(S) = \frac{\operatorname{det}\left(\mathbf{L}_{S}\right)}{\operatorname{det}(\mathbf{L}+\mathbf{I})},
\label{equ-k-dpp:dpp-preliminary}
\vspace{-0.5mm}
\end{equation}
where $\mathbf{L}$ is a real, positive semi-definite kernel matrix indexed by the elements of $S$, and $\mathbf{I}$ is the $M \times M$ identity matrix. 
In this standard DPP setup, a probability is assigned to every subset (including the empty set and the entire $\mathbb{V}$). 
This is not desirable in the practical problems that require explicit control over the number of items selected by the model, such as image search engines that provide a fixed-sized array of results in a page \cite{kulesza2012determinantal}. 
If the model based on a standard DPP gives some probability of suggesting no or every image, this is not ideal. Given this situation, an extension of DPP, \textit{i.e.}, $k$-DPP \cite{kulesza2011k}, has been proposed, which conditions a DPP on the cardinality $k$ of the random set. 
%Through this extension, a DPP-based model can be encouraged to provide the most relevant and diverse subset according to the context of specific practical applications by setting the desired size $k$. 

\begin{figure}
  \centering
  \includegraphics[width=.98\linewidth]{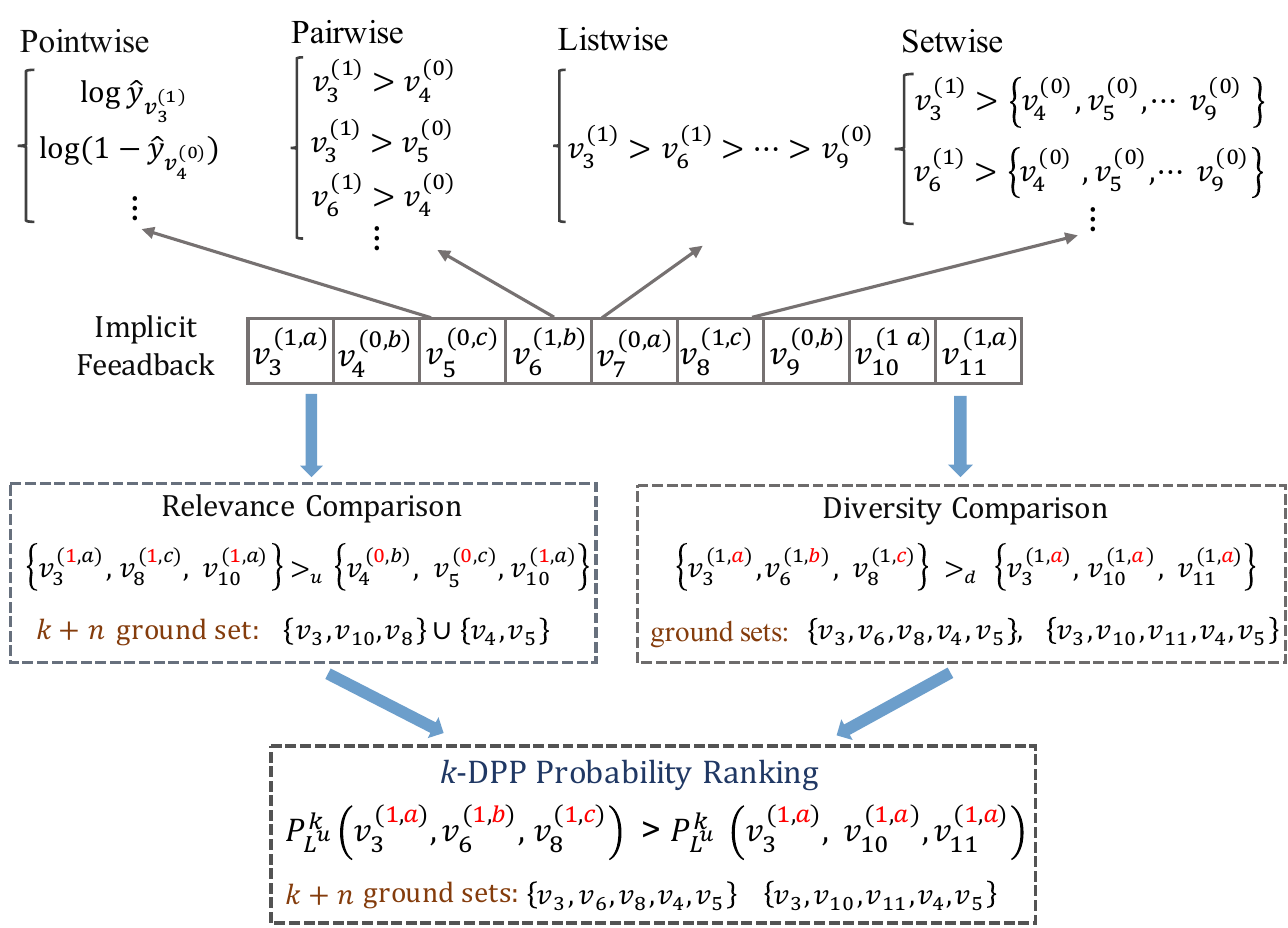}
  \vspace{-0mm}
  \caption{The diagrammatic sketch of L$k$P. Each square in the middle represents an item (denoted as $v$), with the subscript numeral indicating its index. The superscript $1$ ($0$) signifies an observed (unobserved) interaction of user $u$, while $a$, $b$, and $c$ denote three distinct categories of items. The top layer is a simple comparative illustration of different optimization methods. The layers below show our two types (set-level relevance ranking $>_u$ and diversity ranking $>_d$) of comparison, which are integrated using the $k$-DPP probability ranking. The focal comparison points are indicated by red markers.}
  \label{fig-k-dpp:framework}
  \vspace{-5mm}
\end{figure}

\vspace{-1mm}
\subsection{L$k$P Ranking}
\vspace{-1mm}
\label{sec-k-dpp:k-dpp-probability}
%%We now formulate the $k$-DPP probability ranking directed at CF recommendation. A $k$-DPP is a distribution over all subsets $S \subseteq \mathcal{S}$ with cardinality $k$. As a $k$-DPP is obtained by conditioning a standard DPP on the event that the set $S$ has cardinality $k$, constructing a real and symmetric kernel $\mathbf{L}$ remains the primary modelling concern. 
Our objective is to explore $k$-DPP in the context of  optimization criterion, for providing relevant and diverse suggestions. If a purely diversity oriented kernel (whose entries are measurements of pairwise similarity) is adopted directly, the suggested results will bear no relation to user satisfaction, as only the most repulsive items (dissimilar) will be selected.
Therefore, a user-specific kernel $\mathbf{L}^u$ is required, which we obtain via the so-called quality \textit{vs.} diversity decomposition of DPP kernel, 
\vspace{-1mm}
\begin{equation}
\mathbf{L}^u=\operatorname{Diag}\left(\hat{\mathbf{y}}_u\right) \cdot \mathbf{K} \cdot \operatorname{Diag}\left(\hat{\mathbf{y}}_u\right).
\label{equ-k-dpp:Lu}
\vspace{-1mm}
\end{equation}
Following previous studies \cite{warlop2019tensorized, liu2022determinantal}, the $M \times M$ matrix $\mathbf{K}$ models diversity, and is learned from historical interactions.
In this work, diversity is evaluated by category coverage \cite{santos2010exploiting, puthiya2016coverage}, an intuitive and popular diversity-related metric in implicit feedback recommendation. 
To associate $\mathbf{K}$ with the item category, we use diversified item sets (subsets that have a broad coverage) from users' historical interactions as ground truth sets for training \cite{liu2022determinantal, wu2019pd}. In this way, a subset with more categories has a higher possibility of being selected under the DPP as compared to less diverse ones.
To reduce the computational complexity of calculating a $M \times M$ matrix, the diversity kernel is represented using a low-rank factorization $\mathbf{K}=\mathbf{V}^{\top} \mathbf{V}$, learned based on the following objective function, 
\vspace{-1mm}
\begin{equation}
\mathcal{J}\ =\sum_{\left(T^{+}, T^{-}\right) \in \mathcal{T}} \log \operatorname{det}\left(\mathbf{K}_{T^{+}}\right)-\log \operatorname{det}\left(\mathbf{K}_{T^{-}}\right).
\label{equ-k-dpp:objective-function-for-K}
\vspace{-1mm}
\end{equation}
Here, $T^{(+)}$ is an observed diverse set and $T^{(-)}$ represents the set that contains negative items, and $\mathcal T$ denotes the set of paired sets used for training. $\mathbf{K}$ is not related to users, as it is only the diversity-associated kernel. 

The diagonalized $\hat{\mathbf{y}}_u$ plays a personalization role in the kernel $\mathbf{L}^u$, capturing the preference scores of user $u$ (predicted by recommendation models) on entire items. 
The more we can learn from \Cref{equ-k-dpp:Lu} is that the quality component can be represented by different types of relevance calculation, \textit{e.g.}, modeling user-item interaction with inner product \cite{rendle2012bpr, he2020lightgcn} or using nonlinear neural networks \cite{he2017neural}. This extended generality is evaluated in \Cref{sec-k-dpp:experiments}.  

%This means that more generality is showed by formulating $k$-DPP probability for CF models compared to previous setwise ranking approaches that only accept the inner product function (\textit{e.g.}, SetRank and Set2SetRank). 
%We now apply the $k$-DPP to ranking using the $\mathbf{L}^u$ kernel. 
%%As mentioned above, $k$-DPP explicitly controls the number of items selected by the model. Inuitively, the difference between the standard DPP and the $k$-DPP is, in addition to the fixed cardinality, a disparate normalization form compared to Equation 1. Formally, the $k$-DPP gives probabilities directed at ranking optimization,

\subsubsection{Relevance Comparison}
\label{sec-k-dpp:relevance-comparison}
To achieve a set-level relevance comparison by comparing $k$-DPP subsets probabilities, we propose a specialized $k$-DPP distribution dedicated to personalized ranking over a specific $k+n$ ground set (with $k$ observed items in $Y_u^{+}$ and $n$ random items in $Y_u^{-}$). 
It is important to stress that a $k$-DPP assigns probabilities only to subsets of a specific size (cardinality $k$).
That is, given a $k$-DPP distribution over $k+n$ ground set from user $u$, we can only compare probabilities of $k$-sized subsets. 
We opt to define this ground set rather than directly using $\mathbb{V}$, primarily due to considerations of computational efficiency, as learning the $k$-DPP probabilities over $k+n$ items is far more efficient than attempting to do so over the entire item set. 
Another fundamental motivation behind this $k+n$ scheme lies in the necessity for our $k$-DPP distribution to exhibit interpretability in terms of ranking. 
%Item recommendation is inherently a ranking-driven problem, which is also the reason why BPR and listwise optimizations are developed and widely used. 
By defining the $k+n$ ground set, we can confer the tailored $k$-DPP distribution with the \textit{relevance ranking interpretation} of implicitly comparing item sets' relevance rankings. This, in turn, enables the construction of a concrete set-level optimization criterion for personalized ranking.
%The $k+n$ ground set ensures that our $k$-DPP distribution is not only computationally efficient but also geared towards optimizing ranking.

Specifically, given a cardinality $k$ constraint, we consider two types of subsets from the $k+n$ ground set, \textit{i.e.}, the unique \textit{target subset} filled with all $k$ targets and \textit{other $k$-subsets} that are comprised of $k-z$ targets and $z$ unobserved items ($z \in \mathbb{N}^+$ and $z \leq k$, and $z=k$ denotes a subset of unobserved items). 
As shown in the relevance comparison part in \Cref{fig-k-dpp:framework} with 3 ($k$) targets and 2 ($n$) negative samples, it becomes intuitively clear that, given equal-sized sets, we aim to achieve a ranking comparison where the $k$-DPP probability of the sole \textit{target subset} {\fontsize{8}{7}$\small \left\{v_3^{(1, a)}, v_8^{(1, c)}, v_{10}^{(1, a)}\right\}$} is higher than the probabilities of all other $k$-subsets such as {\fontsize{8}{7}$\small \left\{v_4^{(0, b)}, v_5^{(0, c)}, v_{10}^{(1, a)}\right\}$} ($z=2$).

We consider two additional scenarios that help to intuit why the $k+n$ ground set is needed. 
First, if more than $k$ items in $Y_u^{+}$ are added into the ground set, it necessitates the comparison of a $k$-DPP target subset with some subsets that are formed by other combinations of $k$ targets. For instance, in the assumed ground set, it is possible to observe a ranking comparison like {\fontsize{8}{7}$\small \left\{v_3^{(1, a)}, v_8^{(1, c)}, v_{10}^{(1, a)}\right\} >_u \left\{v_3^{(1, a)}, v_{6}^{(1, b)}, v_8^{(1, c)}\right\}$}. 
Second, when applying a standard DPP over the same $k+n$ ground set for personalized ranking, it entails competition between a drawn DPP subset with $k$ targets and all other target subsets of variable cardinality (\textit{e.g.}, one target or two targets), resulting in ranking comparisons such as {\fontsize{7}{7}$\left\{v_3^{(1, a)}, v_8^{(1, c)}, v_{10}^{(1, a)}\right\} >_u \left\{v_3^{(1, a)}, v_8^{(1, c)}\right\}$}.
These will lead to an obscure training signal that does not properly reflect the nature of the set valued targets.
%which implies that the DPP probability of $S$ should be greater than other target subsets, 
% and further explain the intuition of choosing $k+n$ ground set and applying $k$-DPP. 
%As the diversity factor is unified into the personalized kernel $\mathbf{L}_u$, the $k$-DPP probability for ranking is balanced by the diverse kernel $\mathbf{K}$. For example, the ranking probability of a target subset with only one category of items might be diminished in training process by the pre-learned diverse kernel, and thus the corresponding ranking will be affected. 
%We will further explain and validate the ranking implication behind the tailored $k$-DPP distribution in next section.

\subsubsection{Diversity Comparison}
\label{sec-k-dpp:diversity-comparison}
To implement diversity comparison, we also need to consider the $k$-DPP scenario. This consideration allows us to rank diversity in a way that is not affected by differences in set sizes, thus ensuring a fair comparison across subsets. 
Our methodology brings in diversity comparison based on the diverse kernel $\mathbf{K}$. Its learning objective (\Cref{equ-k-dpp:objective-function-for-K}) facilitates attributing larger log-determinant values to sets encompassing a broader range of categories, thereby endowing the $k$-DPP distribution with the \textit{diversity ranking interpretation} (detailed in the next section). As depicted in \Cref{fig-k-dpp:framework}, when focusing solely on the degree of diversity (categories), the corresponding model is expected to favor sets that encompass a greater number of distinct categories, that is, {\fontsize{7}{7}$ \left\{v_3^{(1, a)}, v_6^{(1, b)}, v_8^{(1, c)}\right\}>_d\left\{v_3^{(1, a)}, v_{10}^{(1, a)}, v_{11}^{(1, a)}\right\}$}. 
Since $\mathbf{K}$ does not bear personalization, the focus of comparison is solely on categories. We specifically evaluate the diversity of relevant items, as we aim to guide models towards diversifying the content that is of interest to users, rather than blindly pursuing diversity for its own sake.

\subsubsection{$k$-DPP Comparison}
As discussed in \Cref{sec-k-dpp:relevance-comparison}, 
the tailored $k$-DPP distribution is formulated as
\vspace{-1mm}
\begin{equation}
P_{\mathbf{L}^{(u,k+n)}}^{k}(S_u^{+k})=\frac{\operatorname{det}\left(\mathbf{L}^{(u,k+n)}_{S_u^{+k}}\right)}{\sum_{\left|S^{\prime}\right|=k} \operatorname{det}\left(\mathbf{L}^{(u,k+n)}_{S^{\prime}}\right)}.
\label{equ-k-dpp:L-k+n-positive}
\vspace{-1mm}
\end{equation}
$S_u^{+k}$ represents the $k$-sized \textit{target subset} drawn from the $k+n$ ground set. $\mathbf{L}^{(u,k+n)} \in \mathbb{R}^{(k+n) \times (k+n)}$ indicates the personalized kernel on $k+n$ ground set from user $u$ perspective, which is used to calculate the probability of $S_u^{+k}$ being a $k$-DPP. It is directly calculated by incorporating predicted scores of $k+n$ items and sub-matrix of $\mathbf{K}$ indexed by $k+n$ items referring to \Cref{equ-k-dpp:Lu}.
We use $S^{\prime}$ to denote every cardinality $k$ set from the $k+n$ ground set in the normalization constant. Hence in the tailored $k$-DPP setup a $k$-subset competes only with sets of the same cardinality, while in a standard DPP every $k$-set competes with all other subsets of the ground set. 

For an easier understanding of relevance comparison, we can initially assume that $\mathbf{L}^{(u,k+n)}$ is solely comprised of relevance scores, \textit{i.e.}, $\operatorname{Diag}\left(\hat{\mathbf{y}}_u\right)$. By maximizing the tailored $k$-DPP probability, our optimization criterion L$k$P learns to favor the unique target subset $S_u^{+k}$ over all other subsets of the same size. This implies that the target subset is endowed with a higher probability of being a $k$-DPP-distributed subset than all other $k$-subsets from $k+n$ ground set. This imparts the \textit{relevance ranking interpretation} to the $k$-DPP distribution from the preference perspective of user $u$. 

As mentioned in \Cref{sec-k-dpp:diversity-comparison}, the pre-learned diverse kernel $\mathbf{K}$ (incorporated into $\mathbf{L}^{(u,k+n)}$) enables the tailored $k$-DPP distribution to compare diversity rankings. We can use the following equation to demonstrate the role of $\mathbf{K}$, 
\vspace{-0.8mm}
\begin{equation}
\begin{aligned}
\log P\left(S_u^{+k}\right) 
& \propto \log \operatorname{det}\left(\mathbf{L}^{(u,k+n)}_{S_u^{+k}}\right)\\ &=\sum_{i \in S_u^{+k}} \log \left(\hat{y}_{(u, i)}^2\right)+\log \operatorname{det}\left(\mathbf{K}_{S_u^{+k}}\right).
\label{equ-k-dpp:log-det}
\vspace{-1.2mm}
\end{aligned}
\end{equation}
The log-probability of $S_u^{+k}$ is proportional to the log-determinant that can be further decomposed into two components: the aggregated relevance scores and the determinant of the sub-matrix of $\mathbf{K}$ indexed by $S_u^{+k}$. 
This is applicable when considering two target subsets drawn from different $k$-DPP distributions, each over their unique $k+n$ ground sets sampled from the same user's implicit feedback. As these two target subsets consist of positive items, we can assume their relevance scores to be similar. Therefore, in the comparison of their $k$-DPP probabilities, the decisive factor lies in the diversity factor. Looking at \Cref{equ-k-dpp:objective-function-for-K}, it becomes evident that the learning process for the diverse kernel is geared towards maximizing the determinant value of diverse subset. 
Therefore, in a comparison scenario between two target subsets drawn from respective $k$-DPP distributions, the one exhibiting a broader category coverage tends to possess a higher $k$-DPP probability. 
Consequently, this effect of the \textit{diversity ranking interpretation} influences the unified ($k$-DPP) ranking. 

The concrete set-level optimization criterion L$k$P is thus built upon the tailored $k$-DPP distribution as defined in \Cref{equ-k-dpp:L-k+n-positive}. A target subset characterized by a broader category coverage within the $k$-DPP is inclined to be assigned a higher probability than other monotonous target subsets, as shown in the $k$-DPP probability ranking part of \Cref{fig-k-dpp:framework}, \textit{i.e.}, {\fontsize{8}{7}$P_{\mathbf{L}^u}^k\left(v_3^{(1, a)}, v_6^{(1, b)}, v_8^{(1, c)}\right)>P_{\mathbf{L}^u}^k\left(v_3^{(1, a)}, v_{10}^{(1, a)}, v_{11}^{(1, a)}\right)$}.
This ultimately enables item recommendation models to provide users with not only diverse but also relevant results.

It is important to note that the diverse kernel $\mathbf{K}$ is pre-trained and remains fixed throughout the process of maximizing L$k$P. Its role lies in balancing the probabilities of the drawn target subsets, ensuring a fair and diverse representation in the recommendations. The diversity across $k$-subsets within the same $k$-DPP distribution is not needed to consider. This is because the influence of diversity remains constant in the learning process of the optimization criterion. The objective of maximizing the relevance score of the \textit{target subset} favors it with a higher ranking compared to all \textit{other $k$-subsets}. Therefore, the unified ranking is unaffected by the diversity level of other $k$-subsets in the criterion optimizing.
However, across different $k$-DPP distributions, different target sets are all learned to enhance their relevance scores, thereby making the difference in diversity among these subsets a key determinant of their respective rankings.
%thesis under examination can be improved according to this version

\begin{algorithm}[t]
\caption{Computing elementary symmetric polynomials}\label{euclid}
\small
\linespread{0.95}\selectfont
\vspace{-0.mm}
\begin{algorithmic}
\Statex \textbf{Input:} $k \text {, eigenvalues } \lambda_1, \lambda_2, \ldots \lambda_{(k+n)}$
\Statex \textbf{Output: $e_k\left(\lambda_1, \lambda_2, \ldots, \lambda_{(k+n)}\right)$}
\State $e_0^m \leftarrow 1 \quad \forall \ m \in\{0,1,2, \ldots, k+n\}$
\State $e_l^0 \leftarrow 0 \quad \forall \ l \in\{1,2, \ldots, k\}$
\For{$l=1,2, \ldots k$}
\For{$m=1,2, \ldots, k+n$}
    \State $e_l^m \leftarrow e_l^{m-1}+\lambda_m e_{l-1}^{m-1}$
\EndFor
\EndFor
\Statex \textbf{Return:} $e_k^{(k+n)}$ 
\end{algorithmic}
\vspace{-0.mm}
\end{algorithm}

To sum up, the relevance ranking interpretation signifies that a $k$-DPP inherently captures the implicit ranking comparison between the target subset and other $k$-subsets within a $k$-DPP distribution. 
On the other hand, the diversity ranking interpretation refers to the global balancing of relevance and diversity across different $k$-DPP distributions (over various $k+n$ ground sets), facilitated by the pre-learned diverse kernel.
The relevance and diversity comparisons are respectively demonstrated through two ranking examples in experiments. 

%The fusion of the pre-trained diverse kernel and the relevance scores that need to be predicted creates a $k$-DPP distribution on the balanced personalized kernel, which is inclined to allocate higher probabilities to item sets that are not only diverse but also relevant. 
%Thus, when recommendation models are developed by maximizing the given probabilities (optimization criterion) in \Cref{equ-k-dpp:L-k+n-positive}, 
%It is important to note that $S_u^{+k}$ is the unique target set, since it contains all $k$ targets within the $k+n$ ground set. Conversely, all other $k$-sized subsets would include one or more negative items.
\subsubsection{Optimization Approaches (L$k$P$_{PS}$ and L$k$P$_{NPS}$)}

To implement the implicit comparison of $k$-DPP probabilities between the target subset and other $k$-subsets within a specific ground set, it is crucial to accurately compute the probability of the target subset, which requires proper normalization. 
This normalization constant enables us to associate the target subset and other $k$-subsets within a $k$-DPP distribution, thereby establishing implicit competition relationships, which is  
\vspace{-1mm}
\begin{equation}
Z_k=\sum_{\left|S^{\prime}\right|=k} \operatorname{det}\left(\mathbf{L}^{(u,k+n)}_{S^{\prime}}\right)=e_k\left(\lambda_1, \lambda_2, \ldots, \lambda_{(k+n)}\right). 
\vspace{-1mm}
\end{equation}

Here $\lambda_1, \lambda_2, \ldots, \lambda_{(k+n)}$ are the eigenvalues of $\mathbf{L}^{(u,k+n)}$ \cite{kulesza2012determinantal, gelfand1989lectures}. This formulation has been proved by examining the characteristic polynomial of DPP kernel $\mathbf{L}, \operatorname{det}(\mathbf{L}-\lambda \mathbf{I})$ \cite{fisher1978analysis} or using properties of DPPs \cite{kulesza2012determinantal}.
The recursive algorithm given in Algorithm 1 is used to compute the $k$th elementary symmetric polynomial $e_k$ on $\lambda_1, \lambda_2 \ldots, \lambda_{(k+n)}$, which runs in $O((k+n) k)$ time. %The recursion for $e_k$ is $k$ levels deep, and each level takes $O(k)$ time. 
Thus we can efficiently normalize a tailored $k$-DPP with $n+k$ ground set in $O\left((k+n)k\right)$ time to implicitly carry out the set-level ranking comparison.

%To ensure that the probability of $k$-DPP is compared under the same standard, w
We assign a common $k$ and $n$ for all users. As a shorthand, we will directly use $\mathbf{L}^u$ instead of $\mathbf{L}^{(u,k+n)}$. %%when the meaning is clear. 
Now we can formulate the learning objective to derive our optimization criterion for $k$-DPP probability-based set-level ranking,
\vspace{-1mm}
\begin{equation}
\begin{aligned}
\mathcal{L} =\prod_u   \prod_{S_u^{+k} \in \mathcal{S}_u^{+}}  P_{\mathbf{L}^u}^{k}\left(S_u^{+k}\right) 
= \sum_{u}  \sum_{S_u^{+k} \in \mathcal{S}_u^{+}}  \log \left(P_{\mathbf{L}^u}^{k}\left(S_u^{+k}\right)\right).
\label{equ-k-dpp:objective_p}
\end{aligned}
\vspace{-1mm}
\end{equation}
$\mathcal{S}_u^{+}$ denotes the collection of $k$-sized target sets of user $u$ adopted in the training process. 
In practical experiments, it is not necessary to take all $k$-subsets of $Y_u^+$ into account, given the computation burden and partially redundant calculations. Instead we guarantee that each target item of user $u$ is selected in a $k$-DPP at least once. 
In this way, we ensure that the number of set-level training instances used in our experiments is not greater than the pointwise method or BPR optimization, making for a fair comparison.

\Cref{equ-k-dpp:objective_p} appropriately increases the rank of target subset, but the probability of unobserved item subsets is not explicitly integrated. 
Previous ranking optimization approaches, \textit{e.g.}, BPR, SetRank, and Set2SetRank, usually integrate a ranking pair by providing an observed instance with a randomly selected lower-ranked candidate. Unlike this common approach, we take the unobserved items into account in a probabilistic manner. 
We first formulate the probability of selecting $k$ unobserved items of user $u$ as a $k$-DPP,
\vspace{-1mm}
\begin{equation}
P_{\mathbf{L^u}}^{k}(S_u^{-k})=\frac{\operatorname{det}\left(\mathbf{L}^u_{S_u^{-k}}\right)}{\sum_{\left|S^{\prime}\right|=k} \operatorname{det}\left(\mathbf{L}^{u}_{S^{\prime}}\right)}.
\label{equ-k-dpp:L-k+n-negative}
\vspace{-1mm}
\end{equation}
$S_u^{-k}$ is a negative $k$-set selected from a specific ground set. 
%%That is, the normalization is calculated according to the same algorithm as used in Appendix A.2. 

Our intuition in considering the probability of $S_u^{-k}$ is to widen the gap between targets and unobserved items in a $k$-DPP, \textit{i.e.}, not only increasing the ranking of $S_u^{+k}$ but decreasing that of $S_u^{-k}$. 
To achieve this, we attempt to model an integrated probability that the event that $S_u^{+k}$ is drawn from the specific $k-$DPP, but $S_u^{-k}$ explicitly is not. Basic rules of probability give
\vspace{-1mm}
\begin{equation}
\begin{aligned}
P\left(S_u^{+k} \text{and not } S_u^{-k}\right) & = P\left(S_u^{+k}\right)P\left(\text{not } S_u^{-k} \middle| S_u^{+k}\right) \\ & \approx P\left(S_u^{+k}\right) P\left(\text{not } S_u^{-k}\right), 
\end{aligned}
\vspace{-1mm}
\end{equation}
where $P$ is actually a $k$-DPP on the $k+n$ ground set and the final approximation is an independence assumption. 
Given $P\left(\text{not } S_u^{-k}\right) = 1 - P\left( S_u^{-k}\right)$, we finally obtain the logarithmic objective function,
\vspace{-1mm}
\begin{equation}
\begin{aligned}
\mathcal{L} =&\prod_{(S_u^{+k}, S_u^{-k}) \in \mathcal{S}_u} \left(P_{\mathbf{L}^u}^{k}\left(S_u^{+k}\right)\right)  \left(1-P_{\mathbf{L}^u}^{k}\left(S_u^{-k}\right)\right) \\
&=\sum_{(S_u^{+k}, S_u^{-k}) \in \mathcal{S}_u} \log \left(P_{\mathbf{L}^u}^{k}\left(S_u^{+k}\right)\right) + \log\left(1-P_{\mathbf{L}^u}^{k}\left(S_u^{-k}\right)\right),
\label{equ-k-dpp:objective_np}
\end{aligned}
\vspace{-1mm}
\end{equation}
which can be calculated by combining with \Cref{equ-k-dpp:L-k+n-positive} and \Cref{equ-k-dpp:L-k+n-negative}. $\mathcal{S}_u$ represents the compilation of $k$-sized subsets from the above described ground set. 

The exclusion of $S_u^{-k}$ implies that a $k$-set with unobserved items is supposed to have a lower rank compared to other subsets (with cardinality $k$) that contain one or more targets. 
To avoid extra comparisons between unobserved items, \textit{e.g.}, two $k$-sized subsets with different combinations of unobserved items, we will set $n=k$ for the $k+n$ ground set when explicitly considering exclusion probability. 
We find that this function of considering inclusion of $S_u^{+k}$ and exclusion of $S_u^{-k}$ has a similar formalization to that of the popular binary cross-entropy loss. However, our optimization criterion essentially shows distinct implication behind the formulation (\Cref{equ-k-dpp:objective_np}); specifically, our proposed formulation focuses on set-level comparison.  
In this setting, 

Overall, our optimization criterion focuses on learning the probabilities of $k$-DPPs on specific ground sets for each user to optimize diversity-aware personalized ranking. Two types of optimization approaches are proposed, L$k$P$_{PS}$ and L$k$P$_{NPS}$, wherein only the \textbf{p}ositive sub\textbf{s}et and both exclusion of \textbf{n}egative items (unobserved elements) and inclusion of \textbf{p}ositive sub\textbf{s}et are considered, respectively. 

\vspace{-1mm}
\subsection{L$k$P Optimization}
\vspace{-1mm}
We provide the gradient computation to demonstrate how to maximize our optimization criterion L$k$P.
To learn the parameters $\Theta$ with respect to the model, we can maximize the log-likelihood, derived by substituting \Cref{equ-k-dpp:L-k+n-positive} into \cref{equ-k-dpp:objective_p} and subsequently applying a logarithmic operation,
\vspace{-1mm}
\begin{equation}
\begin{aligned}
\mathcal{L}(\Theta) &= \sum_{S \in D_S, S^{+} \in S} \log \operatorname{det}\left(\mathbf{L}_{S^{+}}(\Theta)\right) \\
&- \sum_{S \in D_S, S^{\prime} \in S} \log \sum_{\left|S^{\prime}\right|=k} \operatorname{det}\left(\mathbf{L}_{S^{\prime}}(\Theta)\right). 
\label{equ-k-dpp:objective-theta}
\end{aligned}
\vspace{-1mm}
\end{equation}
For brevity, $D_S$ represents the collection of training instances of entire users, with $S$ being a specific user's $k+n$ ground set. One thing to note here is that multiple training instances can be selected from each user. To further simplify notations when the meaning is clear, 
%%we directly use $\mathbf{L}$ to denote the DPP kernel specific to a user training instance, rather than using the previously introduced $\mathbf{L}^{(u,k+n)}$; $\mathbf{L}_{S^+}$ represents the sub-matrix of $\mathbf{L}$ indexed by the target set $S^+$ ( represented as $S_u^{+k}$ previously), and $S^{\prime}$ denotes any one of $k$-sized subsets from the ground set. 
we use $\mathbf{L}$ for the DPP kernel specific to a user ground set, bypassing the earlier $\mathbf{L}^{(u,k+n)}$. $\mathbf{L}_{S^+}$ is the sub-matrix of $\mathbf{L}$ indexed by the target set $S^+$ (previously $S_u^{+k}$), and $S^{\prime}$ refers to any $k$-sized subset from the ground set.

Gradient-based methods (\textit{e.g.}, gradient descent and stochastic gradient descent) provide attractive approaches in learning parameters $\Theta$ of DPP kernel because of their theoretical guarantees, but the gradient of the log-likelihood $\mathcal{L}(\Theta)$ is required. In the discrete DPP setting, this gradient can be computed straightforwardly,
\vspace{-1mm}
\begin{equation}
\begin{aligned}
\frac{d \mathcal{L}(\Theta)}{d \ \Theta} &=  \sum_{S \in D_S, \ S^+ \in S} \operatorname{tr}\left(\mathbf{L}_{S^+}(\Theta)^{-1} \frac{d \mathbf{L}_{S^+}(\Theta)}{d \Theta}\right) \\
& -\sum_{S \in D_S, \ S^{\prime} \in S} \sum_{|S^{\prime}|=k } w_{S^{\prime}} \operatorname{tr}\left(\mathbf{L}_{S^{\prime}}(\Theta)^{-1} \frac{d \mathbf{L}_{S^{\prime}}(\Theta)}{d \Theta}\right),
\label{equ-k-dpp:gradient-theta}
\end{aligned}
\vspace{-1mm}
\end{equation}
where $w_{S^{\prime}}$ represents the importance (normalized probability) of the $k$-sized subset $S^{\prime}$ in the ground set.
%, calculated as \begin{equation} w_{S^{\prime}} = \frac{\operatorname{det}\left(\mathbf{L}_{S^{\prime}}(\Theta)\right)}{\sum_{|{S^{\prime}}|=k} \operatorname{det}\left(\mathbf{L}_{S^{\prime}}(\Theta)\right)} \label{equ-k-dpp:gradient_ws} \end{equation}

%%As we employ quality \textit{vs.} diversity decomposition to balance DPP kernel, 
The entry of kernel $\mathbf{L}$ can be written as,  
\vspace{-1mm}
\begin{equation}
\mathbf{L}_{i j}=\exp{\left(\mathbf{e}_u\mathbf{e}_i^{\top}\right)}\mathbf{K}_{ij}\exp{\left(\mathbf{e}_u\mathbf{e}_j^{\top}\right)},
\label{equ-k-dpp:gradient_lij}
\vspace{-1mm}
\end{equation}
where $\mathbf{K}$ is the pre-learned diverse kernel.  
The quality value $\exp{\left(\mathbf{e}_u\mathbf{e}_i^{\top}\right)}$ corresponds the relevance of item $i$ to user $u$, calculated based on the user and item embeddings, which are exactly the parameters that we need to learn. 

%Specifically, we can compute the gradient of DPP kernel with respect to the parameter $e_u^{(d)}$ ($d$ is the dimension number of user $u$'s feature vector) as
%\begin{equation}
%\begin{aligned}
%\frac{\partial \mathbf{L}_{i j}}{\partial e_u^{(d)}} &=\exp \left(\mathbf{e}_u \mathbf{e}_i^{\top}\right) \mathbf{K}_{i j} \exp \left(\mathbf{e}_u \mathbf{e}_j^{\top}\right) (e_i^{(d)}+ e_j^{(d)}). \\
%&=\mathbf{L}_{i j}\left(e_i^{(d)} + e_j^{(d)}\right).
%\label{equ-k-dpp:gradient_eu}
%\end{aligned}
%\end{equation}

Denote $\mathbf{R}_{ij} = \mathbf{L}_{i j}\left(e_i^d+e_j^d \right)$, we can obtain the gradient of DPP kernel with respect to the parameter $e_u^{(d)}$ ($d$ is the dimension number of user $u$'s feature vector) as
\vspace{-1mm}
\begin{equation}
\begin{aligned}
\frac{d \mathcal{L}(\Theta)}{d e_u^d} &=  \sum_{S \in D_S, \ S^+ \in S} \operatorname{tr}\left(\mathbf{L}_{S^+}(\Theta)^{-1} \mathbf{R}_{S^+}^{(d)} \right) \\
& -\sum_{S \in D_S, \ S^{\prime} \in S} \sum_{|S^{\prime}|=k } w_{S^{\prime}} \operatorname{tr}\left(\mathbf{L}_{S^{\prime}}(\Theta)^{-1} \mathbf{R}_{S^{\prime}}^{(d)}\right),
\label{equ-k-dpp:gradient_R}
\end{aligned}
\vspace{-1mm}
\end{equation}

%Similarly, we can compute the gradient of DPP kernel entry $\mathbf{L}_{i j}$ with respect to the representation of an item (\textit{i.e.}, $\mathbf{e}_i$ for item $i$) as 
%\begin{equation}
%\begin{aligned}
%\frac{\partial \mathbf{L}_{i j}}{\partial e_i^{(d)}} =
%&=\mathbf{L}_{i j} e_u^{(d)}.
%\label{equ-k-dpp:gradient_ei}
%\end{aligned}
%\end{equation}

Similarly, let $\mathbf{G}_{i j}^{(d)} = \mathbf{L}_{i j}^{(S)}e_u^d$, and then the gradient of $\mathbf{L}_{i j}$ with respect to the representation of an item (\textit{i.e.}, $\mathbf{e}_i$ for item $i$) can be computed by
\vspace{-1mm}
\begin{equation}
\begin{aligned}
\frac{d \mathcal{L}(\Theta)}{d e_i^d} &=  \sum_{S \in D_S, \ S^+ \in S} \operatorname{tr}\left(\mathbf{L}^{(S)}_{S^+}(\Theta)^{-1} \mathbf{G}_{S^+}^{(d)} \right) \\
& -\sum_{S \in D_S, \ S^{\prime} \in S} \sum_{|S^{\prime}|=k } w_{S^{\prime}} \operatorname{tr}\left(\mathbf{L}^{(S)}_{S^{\prime}}(\Theta)^{-1} \mathbf{G}_{S^{\prime}}^{(d)}\right),
\label{equ-k-dpp:gradient_G}
\end{aligned}
\vspace{-1mm}
\end{equation}

%Substitute these two equations (15 and 16) back into the previous gradient formula, and further get the corresponding final results.
From above formulations, it can be demonstrated that our optimization criterion based on set-level DPP probability comparison is differentiable. We therefore can apply the stochastic gradient descent based algorithms with learning rate $\eta$ to optimize the model parameters based on sampled $(u, S) \in D_S$.
In our experiments, we employ Adam and grid search $\eta$ for our approaches and baselines.
%\begin{equation}
%\Theta_i=\Theta_{i-1}-\eta \frac{\partial l(\Theta)}{\partial \Theta}
%\label{equ-k-dpp:gradient_ascent}
%\end{equation}
The gradient calculation process for \Cref{equ-k-dpp:objective_np} is not presented, as \Cref{equ-k-dpp:objective_np} and \Cref{equ-k-dpp:objective_p} share the same basic components and parameters. 

%%implicit meaning, k sized set of targets is supposed to rank higher than all the other sets with k items. For example, a set with 2 positive items and three items. But if we not control the set size, it is not reasonable, for example 5 positive rank higher than 2 positive? ranking consideration is important for cf, ranking optimization problem.This also the reason why we need normalization. endow ranking implication.
%%as the DPP distribution is under cardinality $k$ constraint. 
%the normalization considers all k-sized subsets
%not over all items: hard to explain the ranking, too computationally expensive;

\vspace{-1mm}
\section{Experiments}
\vspace{-1mm}
\label{sec-k-dpp:experiments}
%We perform experiments on three real-world datasets. 
To comprehensively demonstrate the superiority of our approaches, three levels of evaluations and comparisons are conducted: \textit{i}) six L$k$P based variants are proposed and evaluated; \textit{ii}) two types of implementation of L$k$P for recommendation are achieved based on Matrix Factorization (MF) and Graph Convolutional Network (GCN), and are compared with state-of-the-art baselines; \textit{iii}) through verifying the improvement after applying our approaches on different recent CF methods (reworked models) over their original framework, the generality and availability can be validated.

\vspace{-1mm}
\subsection{Settings}
\vspace{-1mm}

\begin{table}
\centering
  \fontsize{7.7}{8.9}\selectfont
  \caption{Statistics of the datasets.}
  \setlength{\tabcolsep}{3mm}{
  \vspace{-1.2mm}
  \label{tab:statistics}
    \begin{tabular}{ccccc}
    \hline
    \ \ \ Dataset \ &\#Users&\#Items&\#Interactions&\#Categories \ \ \ \\
    \hline
    \textbf{Beauty}& 52.0k & 57.2k & 0.4M & 213\\
    \textbf{ML}& 6.0k & 3.4k & 1.0M & 18\\
    \textbf{Anime}& 73.5k & 12.2k & 1.0M & 43\\
    \hline
    \end{tabular}}
    \label{tab-k-dpp:datasets}
    \vspace{-4mm}
\end{table}

\subsubsection{Datasets} 
%%For evaluating our proposed $k$-DPP based personalized ranking approaches (L$k$P$_{PS}$ and L$k$P$_{NPS}$), 
We select three common and real-world datasets as appropriate, whose category numbers and matrix densities of implicit interactions differ significantly. 
Amazon-review %\footnote{\url{http://jmcauley.ucsd.edu/data/amazon/}} 
contains a series of datasets \cite{he2016ups}, comprised of a large corpus of product ratings. We select \textbf{Beauty} products from the collection, which have 213 categories, respectively. 
MovieLens %\footnote{\url{https://grouplens.org/datasets/movielens/1m/}}  
is a widely used benchmark movie rating dataset. We use the version \textbf{ML}-1M that includes one million user ratings to 18 categories of movies.
\textbf{Anime}
%\footnote{\url{https://www.kaggle.com/CooperUnion/anime-recommendations-database}} 
dataset consists of user ratings to anime from \texttt{myanimelist.net}. 
All items in Anime are grouped into 43 categories.
We transform these datasets into implicit data, where each entry is marked as 1/0, depending on whether the ratings are 5.
We filter out long-tailed users and items with
fewer than 10 interactions for all datasets.
For each user, we randomly select 20\% of the rated items as ground truth for testing, and 70\% and 10\% ratings constitute the training and validation set, respectively.
The statistics of the three datasets are summarized in \Cref{tab-k-dpp:datasets}. We can note that Beauty have more plentiful product categories and lower densities compared to ML and Anime. 

\subsubsection{Baselines and Metrics} We place all compared methods into three groups according to different levels of comparisons. \\
\textbf{Six variants} derived from $k$-DPP probability are given, \textit{i.e.}, PR, PS, NPR, NPS, PSE, and NPSE. In these methods, P means the usage of the target set inclusion probability, and N denotes the consideration of the exclusion of unobserved items. 
S represents a method of sampling for a training instance, \textit{i.e}, selecting of $k$ observed items in the order they occurred using a sliding window, along with $n$ randomly selected unobserved items.
R signifies random selection, \textit{i.e.},  randomly selecting $k+n$ items ($k$ targets and $n$ un-interacted items) from user's $1/0$ feedback. 
Inspired by the original definition of entries of DPP diversity kernel (\textit{i.e.}, measurements of similarity), we propose another type of formulation for diversity kernel $\mathbf{K}$. 
It directly regards any two items’ embeddings as their feature vectors, which are used to calculate the similarity of item pairs following the calculation manner of Gaussian kernel \cite{affandi2014learning, yu2009recommendation}. 
%In this setting, diversity kernel $\mathbf{K}$ is associated with the diversity measurement of intra-list distance (ILD) \cite{puthiya2016coverage,zhang2008avoiding}. 
E denotes this new formulation of diversity factor, instead of pre-learned $\mathbf{K}$. 
In the E setting, the optimization process incorporates the diversity factor directly. The intention is to learn item representations in a direction that promotes dissimilarity.
For example, NPSE represents a method that uses \Cref{equ-k-dpp:objective_np} (exclusion of \textbf{n}egative items and inclusion of \textbf{p}ositive items) as the objective function trained using \textbf{s}equentially selected target sets, and calculates the diverse kernel based on trainable \textbf{e}mbeddings. 
\\ \textbf{Four state-of-the-art and popular objective functions} are included: binary cross-entropy (BCE) \cite{he2017neural}, BPR \cite{rendle2012bpr}, SetRank \cite{wang2020setrank}, and S2SRank (\textit{i.e.}, Set2SetRank) \cite{chen2021set2setrank}. All optimization baselines and our approaches are deployed in scenarios of MF- and GCN-based collaborative filtering (CF) recommendation models for a thorough evaluation. 
Our choice to implement all ranking optimization approaches using basic MF- and GCN-based CF models stems from a thoughtful rationale. As we mentioned in \Cref{sec-k-dpp:related-work}, MF is the most common CF technique and is widely adopted as a baseline in recommendation research due to its simplicity and effectiveness in capturing latent factors and item-user interactions. GCN \cite{he2020lightgcn, wang2019neural, chen2022ba} has gained substantial prominence in the field of collaborative deep learning due to their ability to leverage graph structures and capture intricate item-item relationships. 
By selecting both MF- and GCN-based CF models for validation, we ensure a comprehensive evaluation across a spectrum of traditional and state-of-the-art approaches in item recommendation. 
\\  \textbf{Two distinct seminal CF models} in item recommendation field, \textit{i.e.}, GCMC \cite{berg2017graph} and NeuMF \cite{he2017neural}, are selected to evaluate the generality and applicability of our concrete set-level ranking optimization approaches. To take different types of CF methods into account as comprehensively as possible, we select two seminal and distinct CF models as baselines in this part of the evaluation. GCMC \cite{berg2017graph} is a seminal graph-based auto-encoder work for recommender systems, which inspired many graph neural networks based studies in the CF field, \textit{e.g.}, NGCF \cite{he2017neural}, lightGCN \cite{he2020lightgcn}, and KGAT \cite{wang2019kgat}. 
It applies negative log likelihood as loss, and a probability distribution over possible rating levels by a softmax function is produced, which is distinct from commonly used relevance calculation methods (\textit{i.e.}, inner product calculation and nonlinear neural networks introduced in \Cref{sec-k-dpp:k-dpp-probability}). 
NeuMF \cite{he2017neural} is a classic and comprehensive model that combines GMF with MLP, which provides inspiration to numerous related studies \cite{zhang2021group, yin2019social, zheng2020price}. Its relevance is predicted through classifying the interaction from a user to a specific item through nonlinear neural networks. This model is optimized based on classification-aware binary cross-entropy loss.
%%To take different types of CF methods into account as comprehensively as possible, we select two seminal and distinct CF models as baselines in this part of the evaluation. 
%%GCMC \cite{berg2017graph} is a seminal graph-based auto-encoder work for recommender systems, which inspired many graph neural networks based studies in the CF field, \textit{e.g.}, NGCF \cite{he2017neural}, lightGCN \cite{he2020lightgcn}, and KGAT \cite{wang2019kgat}. 
%%It applies negative log likelihood as loss, and a probability distribution over possible rating levels by a softmax function is produced, which is distinct from commonly used relevance calculation methods (\textit{i.e.}, inner product calculation and nonlinear neural networks introduced in \Cref{sec-k-dpp:k-dpp-probability}). 
%%NeuMF \cite{he2017neural} is a classic and comprehensive model that combines GMF with MLP, which provides inspiration to numerous related studies \cite{zhang2021group, yin2019social, zheng2020price}. Its relevance is predicted through classifying the interaction from a user to a specific item through nonlinear neural networks. This model is optimized based on classification-aware binary cross-entropy loss. 

Two types of accuracy related metrics, \textit{i.e.}, NDCG@N (Nd) and Recall@N (Re), the popular and intuitive diversity metric --- \textit{Category Coverage} (CC) \cite{wu2019pd, liu2022determinantal, puthiya2016coverage}, and a harmonic F-score (F) \cite{cheng2017learning, liang2021recommending} metric between quality (accuracy) and diversity are adopted. 
As we focus on implicit feedback and there is no explicit feature under consideration, another commonly used diversity metric, \textit{i.e.}, intra-list distanc (ILD), \cite{puthiya2016coverage,zhang2008avoiding} is not adopted in evaluation. 
We set $\mathrm{N} \in\{5,10,20\}$ for comprehensive evaluation. 

\subsubsection{Implementations} For the above baselines, we have carefully explored the corresponding parameters, \textit{i.e.}, the number of dimensions, the learning rate and regularization parameters. 
For a fair comparison we set the trainable embedding vectors of all compared models in this work to the same size (64), and report the best results of each model by tuning the hyperparameters on a validation set. 
For the CF baselines, the reworked models and original approaches both use the default parameters provided by their official source codes. 
A prominent variant of stochastic gradient descent method, Adam, is applied. Experiments are implemented on a NVIDIA Quadro P2000 GPU.
%Our codes will be released upon publication.
\vspace{-1.mm}
\subsection{Comparisons and Analysis}
\vspace{-1.2mm}
%%\multicolumn{2}{c|} control the | in multi columns
%% P:Y+ N:Y^- R: randomly sample S:seq_sample G:gaus kernel or pre-learned
%%max vs. max means our best variant (best F average three values, with * after the name) vs. best other losses
%%denote the %improvement of our non-best variant over best others using *
\begin{table*}[tp]
\centering
  \fontsize{7.4}{9.7}\selectfont
  \caption{Performance comparison between L$k$P and state-of-the-art objective functions, deployed on deep neural network.}
   \vspace{-0.5mm}
  \setlength{\tabcolsep}{1.45mm}{
  \label{tab:performance_comparison}
    \begin{tabular}{lllllllllllllll}
    \Xhline{0.8pt}
    Dataset&\multicolumn{2}{c}{Method}&  Re@5&Re@10&Re@20&Nd@5&Nd@10&Nd@20&CC@5&CC@10&CC@20&F@5&F@10&F@20  \cr \Xhline{0.6pt}
    \multirow{12}{*}[1pt]{Beauty} &  \multirow{6}{*}{L$k$P}
    &PR &0.0788&0.1235&0.1754&0.0808&0.1006&0.1209&0.0579&0.0891&\uwave{0.1312}&0.0671&0.0992& 0.1391 \\ 
    &&PS &0.0806&0.1281&0.1841&0.0794&0.1001&0.1213&\uwave{0.0575}&\uwave{0.0886}&0.1360&0.0669& 0.0998& 0.1439\\
    &&NPR &0.0820&0.1256&0.1756&0.0801&0.0989&0.1184&0.0588&0.0914&0.1354&0.0682& 0.1008& 0.1410 \\ 
    &&NPS &\textbf{0.0868}&\textbf{0.1345}&\textbf{0.1874}&\textbf{0.0878}&\textbf{0.1081}&\textbf{0.1283}&0.0578&0.0894&0.1334&\textbf{0.0696}&\textbf{0.1029}&\textbf{0.1446} \\
    &&PSE &0.0726&0.1163&0.1717&0.0716&0.0908&0.1121&0.0589&0.0921& 0.1377& 0.0648&0.0975&0.1398 \\
    &&NPSE &\uwave{0.0684}&\uwave{0.1078}&\uwave{0.1596}&\uwave{0.0695}&\uwave{0.0871}&\uwave{0.1068}&\textbf{0.0604}&\textbf{0.0950}&\textbf{0.1446}&\uwave{0.0644}&\uwave{0.0962}& \uwave{0.1387} \\
    \cline{2-15}
    &  \multirow{4}{*}[1pt]{Others}
    &BPR &0.0694&0.1097\boldmath{$^{-}$}&0.1544\boldmath{$^{-}$}&0.0704&0.0876&0.1049\boldmath{$^{-}$}&0.0569&0.0879&0.1301&0.0627& 0.0930\boldmath{$^{-}$}& 0.1299\boldmath{$^{-}$} \\
    &&BCE &0.0672\boldmath{$^{-}$}&0.1158&0.1700&0.0656\boldmath{$^{-}$}&0.0865\boldmath{$^{-}$}&0.1070&0.0568&0.0883&0.1336\boldmath{$^{+}$}&0.0612\boldmath{$^{-}$}& 0.0943& 0.1360 \\
    &&SetRank &0.0775&0.1228&0.1746&0.0772&0.0996&0.1166&0.0574\boldmath{$^{+}$}&0.0884\boldmath{$^{+}$}&0.1278\boldmath{$^{-}$}&0.0659\boldmath{$^{+}$}& 0.0985\boldmath{$^{+}$}& 0.1361\\
    &&S2SRank &0.0783\boldmath{$^{+}$}&0.1240\boldmath{$^{+}$}&0.1752\boldmath{$^{+}$}&0.0776\boldmath{$^{+}$}&0.1004\boldmath{$^{+}$}&0.1172\boldmath{$^{+}$}&0.0566\boldmath{$^{-}$}&0.0875\boldmath{$^{-}$}&0.1303&0.0656& 0.0983& 0.1378\boldmath{$^{+}$} \\
    \cline{2-15}
    &\multicolumn{2}{c}{\textit{max vs. min} (\%)}&29.17&22.61&21.37&33.84&24.97&22.31&6.71&8.57&13.14&13.59& 10.72& 11.34 \\
    &\multicolumn{2}{c}{\textit{max vs. max} (\%)}&10.86&8.47&6.96&13.14&7.67&9.47&5.23&7.47&8.23& 5.54& 4.51&4.94\\
    \Xhline{0.6pt}
    \multirow{12}{*}[1pt]{ML} &  \multirow{6}{*}{L$k$P}
    &PR&0.0831 &0.1276 &0.1896 &0.0895 &0.1082 &0.1281 &0.3417 &0.5270 &0.6582 & 0.1378 & 0.1927 & 0.2559 \\ 
    &&PS&\textbf{0.0869} &0.1354 &0.1957 &\textbf{0.0952} &0.1184 &0.1385 &0.3346 &0.5260 &0.6559 &\textbf{0.1431} & 0.2045 & 0.2663 \\ 
    &&NPR&0.0857 &0.1260 &0.1941 &0.0913 &0.1126 &0.1368 &0.3341 &0.5237 &0.6590 &0.1399 & 0.1943 & 0.2645 \\  
    &&NPS&0.0862 &\textbf{0.1412} &\textbf{0.1972} &0.0950 &\textbf{0.1237} &\textbf{0.1403} &0.3326 &0.5230 &0.6594 &0.1424 & \textbf{0.2114} & \textbf{0.2687} \\ 
    &&PSE&0.0820 &0.1253 &0.1916 &0.0872 &0.1115 &0.1294 &0.3364 &0.5306 &0.6608 &0.1352 & 0.1936 & 0.2583  \\ 
    &&NPSE&\uwave{0.0766} &\uwave{0.1138} &\uwave{0.1754} &\uwave{0.0836} &\uwave{0.0982} &\uwave{0.1260} &\textbf{0.3380} &\textbf{0.5316} &\textbf{0.6615} &\uwave{0.1295} & \uwave{0.1768} & \uwave{0.2456}\\ 
    \cline{2-15}
    &  \multirow{4}{*}[1pt]{Others}
    &BPR&0.0785 &0.1205 &0.1800\boldmath{$^{-}$} &0.0854 &0.1017 &0.1254\boldmath{$^{-}$} &0.3329 &0.5245\boldmath{$^{+}$} &0.6571\boldmath{$^{+}$} &0.1315 & 0.1834 & 0.2478\boldmath{$^{-}$} \\ 
    &&BCE&0.0762\boldmath{$^{-}$} &0.1185\boldmath{$^{-}$} &0.1829 &0.0825\boldmath{$^{-}$} &0.1002\boldmath{$^{-}$} &0.1276 &0.3210\boldmath{$^{-}$} &0.5234 &0.6560 &0.1272\boldmath{$^{-}$} & 0.1809\boldmath{$^{-}$} & 0.2511 \\ 
    &&SetRank&0.0805\boldmath{$^{+}$} &0.1264 &0.1840 &0.0890\boldmath{$^{+}$} &0.1174 &0.1269 &0.3346\boldmath{$^{+}$} &0.5225 &0.6497\boldmath{$^{-}$} &0.1352\boldmath{$^{+}$} & 0.1931 & 0.2509 \\ 
    &&S2SRank&0.0791 &0.1298\boldmath{$^{+}$} &0.1857\boldmath{$^{+}$} &0.0878 &0.1153\boldmath{$^{+}$} &0.1310\boldmath{$^{+}$} &0.3216 &0.5176\boldmath{$^{-}$} &0.6519 &0.1325 & 0.1995\boldmath{$^{+}$} & 0.2548\boldmath{$^{+}$}\\ 
    \cline{2-15}
    &\multicolumn{2}{c}{\textit{max vs. min} (\%)}&14.04 & 19.16 & 9.56 & 15.39 & 23.45 & 11.88 & 6.45 & 2.70 & 2.12 & 12.50 & 16.84 & 8.44 \\ 
    &\multicolumn{2}{c}{\textit{max vs. max} (\%)}&7.95 & 8.78 & 6.19 & 6.97 & 5.46 & 7.10 & 2.12 & 1.35 & 0.97 & 5.84 & 5.96 & 5.46 \\ 
    \Xhline{0.6pt}
    \multirow{11}{*}[1pt]{Anime} &  \multirow{6}{*}{L$k$P}
    &PR &0.0863&0.1403&0.2231&0.1421&0.1449&0.1836&0.3319&0.4538&0.5837&0.1699& 0.2170& 0.3016 \\ 
    &&PS &\textbf{0.0975}&0.1558&0.2442&0.1560&0.1682&0.2043&0.3359&0.4520&0.5825&\textbf{0.1841}&0.2385&0.3238\\
    &&NPR &0.0920&0.1508&0.2340&0.1490&0.1543&0.1932&0.3300&\uwave{0.4517}&\uwave{0.5657}&0.1765&0.2281&0.3101 \\ 
    &&NPS &0.0974&\textbf{0.1560}&\textbf{0.2447}&\textbf{0.1579}&\textbf{0.1692}&\textbf{0.2059}&\uwave{0.3293}&0.4524&0.5758&0.1840& \textbf{0.2392}& \textbf{0.3239} \\
    &&PSE &0.0895&0.1498&0.2388&0.1494&0.1548&0.1966&0.3353&\textbf{0.4662}&0.5949&0.1761& 0.2296& 0.3188 \\
    &&NPSE &\uwave{0.0784}&\uwave{0.1365}&\uwave{0.2089}&\uwave{0.1328}&\uwave{0.1375}&\uwave{0.1762}&\textbf{0.3381}&0.4641&\textbf{0.5953}&\uwave{0.1609}& \uwave{0.2116}& \uwave{0.2910} \\
    \cline{2-15}
    &  \multirow{4}{*}[1pt]{Others}
    &BPR &0.0879&0.1432&0.2184\boldmath{$^{-}$}&0.1443&0.1473&0.1825\boldmath{$^{-}$}&0.3339&0.4519&0.5890\boldmath{$^{+}$}&0.1723& 0.2198& 0.2991\boldmath{$^{-}$} \\
    &&BCE &0.0863\boldmath{$^{-}$}&0.1403\boldmath{$^{-}$}&0.2231&0.1421\boldmath{$^{-}$}&0.1448\boldmath{$^{-}$}&0.1836&0.3320&0.4513\boldmath{$^{-}$}&0.5737&0.1699\boldmath{$^{-}$}& 0.2167\boldmath{$^{-}$}& 0.3003 \\
    &&SetRank &0.0890&0.1446&0.2269&0.1467&0.1501&0.1884&0.3485\boldmath{$^{+}$}&0.4657\boldmath{$^{+}$}&0.5735\boldmath{$^{-}$}&0.1761\boldmath{$^{+}$}& 0.2239& 0.3049 \\
    &&S2SRank &0.0902\boldmath{$^{+}$}&0.1471\boldmath{$^{+}$}&0.2295\boldmath{$^{+}$}&0.1477\boldmath{$^{+}$}&0.1546\boldmath{$^{+}$}&0.1892\boldmath{$^{+}$}&0.3254\boldmath{$^{-}$}&0.4600&0.5793&0.1742& 0.2272\boldmath{$^{+}$}& 0.3076\boldmath{$^{+}$} \\
    \cline{2-15}
    &\multicolumn{2}{c}{\textit{max vs. min} (\%)}&12.98&11.19&12.04&11.12&16.85&12.82&3.90&3.30&3.80&8.30& 10.41& 8.28 \\
    &\multicolumn{2}{c}{\textit{max vs. max} (\%)}&8.09&6.05&6.62&6.91&9.44&8.83&-2.98&0.11&1.07&4.49& 5.29& 5.32 \\
    \Xhline{0.8pt}
    \end{tabular}}
    \label{tab-k-dpp:GCN}
    \vspace{-3mm}
\end{table*}

\subsubsection{Comparisons}
The overall performance comparison 
%between our $k$-DPP based ranking (six variants) and the state-of-the-art objective functions (binary cross-entropy loss and three types of ranking optimization) on three datasets
is reported in \Cref{tab-k-dpp:GCN}. All results are obtained by implementing GCN-based CF model, under the setting of $k=n=5$. 
To maintain the generality of L$k$P, we employ the basic GCN framework that learns representations from high-order connectivities referring to NGCF \cite{wang2019neural}. 
The bold value represents the best result of all methods. The annotation \uwave{\ \ \ \ } denotes the worst performance from L$k$P variants. The superscripts $^+$ and $^-$ label the best and worst performances out of baselines. 
Two types of improvements are calculated, the best performance of our variants \textit{vs.} the best (\textit{max}) and worst (\textit{min}) performances from baselines. 
Three groups of key observations are made: 
 \\
(\textit{i}) \textit{comparison between six} L$k$P \textit{variants}.
\begin{itemize} [leftmargin=*]
\vspace{-0.7mm}
\item Comparing R and S (two types of training instance construction), \textit{i.e.}, PR \textit{vs.} PS and NPR \textit{vs.} NPS, shows that selecting $k$ targets in a certain sequence (S) outperforms the random selection mode (R) \textit{w.r.t.} quality metrics (NDCG and Recall) on three datasets. 
This may be caused by the dependency among items in the target subset, as the items in sequence have clearer correlations (\textit{e.g.}, similar attributes, or the same category) compared to randomly constructed training targets. These correlations naturally captured by $k$-DPP in S training instances contribute to collaborative filtering.
With regard to diversity results, R mode achieves better performance than S, especially on Beauty and ML datasets. This is expected, as randomly selected target subsets are endowed with more diverse correlations than is the case for S. 
In addition, with the same size of training instances, L$k$P using R needs more epochs to achieve convergence compared to S, as R makes it hard to learn the $k$-DPP distribution due to increased randomness. 
These results verify that L$k$P compares (relevance and diversity) rankings at set level and considers correlations within items. 
\item Among the two diversity factor construction methods, \textit{i.e.}, the pre-learned kernel $\mathbf{K}$ using \Cref{equ-k-dpp:objective-function-for-K} (default type) and the item embeddings' similarity kernel (E type), the default type approaches (\textit{i.e.}, PR, PS, NPR, and NPS) achieve better relevance performance than E type methods (\textit{i.e.}, PSE and NPSE). 
However, in general the diversity performance of these two types of kernels is exactly the opposite, especially in the cases of CC@10 and CC@20.  
The main reason behind these results is that the diversity is provided by the pre-learned diverse kernel in the default type, making it relatively easy to balance the quality and diversity. However, E type considers the diversity factor based on the intuition of expanding intra-list distance (\textit{i.e.}, ILD) and tries to directly learn embeddings towards two directions (quality and diversity) by regarding embeddings as feature representations of items. 
The weak overall performance (F) of E type indicates that explicitly learning item embeddings with two objectives finally places the recommendation system in a dilemma, and degrades performance. 
The noteworthy diversity results of E type further indicate that L$k$P facilitates capturing diverse correlations not only from pre-widened category coverage (default) but also from widening ILD (E). 
In addition, the quality results of E type are comparable to those of the baselines (even superior to the baselines on Beauty \textit{w.r.t.} Re@20) demonstrating the superiority of our approaches.
The comparison between R and S above have shown that S mode is more suitable for L$k$P, which is the reason why only the combination of S and E are listed in \Cref{tab-k-dpp:GCN}. 
\item Two main L$k$P based approaches are compared, \textit{i.e}, considering only the probability of selecting a target subset (P) and considering both the inclusion of targets and exclusion of unobserved items (NP), which are based on \Cref{equ-k-dpp:objective_p} and \Cref{equ-k-dpp:objective_np}, respectively. We only compare PS with NPS here, as other variants, \textit{e.g.}, R and E, are not effective enough. 
In general, NPS performs better than PS in both quality and diversity metrics. This suggests that explicitly taking the exclusion probability of unobserved items into account enriches the relevance ranking of L$k$P, and allows the model to capture more relationships between relevance and diversity, thereby preventing the recommendation model from suggesting monotonous results. 
In some cases, \textit{e.g.}, Re@5 and Nd@5 on ML and Re@5 on Anime, PS beats NPS. This may occur because only focusing on selecting preferred items as a $k$-DPP highlights the relevance of potential positive items in the top rankings without the distraction of unobserved items. 
The notable performance of PS and NPS shows the prospect of L$k$P. 
\item We can find that NPSE is generally weaker than PSE in relevance metrics, which is contrary to the previous comparison between P and NP. 
This may arise from the usage of the E type diversity factor, as adding the probabilistic learning of unobserved items in the objective function (\Cref{equ-k-dpp:objective_np}) makes the embedding learning process that balances quality and diversity less congruent, compared to only drawing targets (\Cref{equ-k-dpp:objective_p}).  
\vspace{-0.2mm}
\end{itemize}
(\textit{ii}) \textit{comparison between our approaches and baselines}.
\vspace{-0.2mm}
\begin{itemize} [leftmargin=*]
\item A significant improvement over pointwise loss (BCE) and pairwise ranking (BPR) achieved by our main L$k$P approaches (PS and NPS) is apparent in quality metrics, exceeding 20\% on Beauty, and around 10\% on ML and Anime. 
This improvement trend also exists between L$k$P and other set-level ranking methods (SetRank and S2SRank), \textit{i.e.}, around 10\% on Beauty, and more than 5\% on ML and Anime.
This indicates that L$k$P approaches have clear advantages in dealing with more sparse interaction matrices (Beauty) over baselines, as Beauty has the highest sparsity matrix (as calculated from \Cref{tab-k-dpp:datasets}).
The superiority of our approaches is verified with a significant boost even in comparison with the recently proposed set ranking methods. 
\item Regarding diversity evaluation, PS and NPS can provide more diversified suggestions for users than baselines in most cases. This means that the proposed approaches based on $k$-DPP really consider diversity in the $k$-DPP learning process. 
\item In the comparison of trade-off metric, our approaches outperform baselines on three datasets at different Top-N metrics (5, 10, 20). 
That is, in line with their inherent modeling of both quality and diversity, $k$-DPP based optimization approaches are better at balancing relevance and diversity, compared to common quality \textit{vs.} diversity trade-off methods that sacrifice quality for diversity, \textit{e.g.}, SetRank on Beauty and BCE on ML.
\vspace{-0.3mm}
\end{itemize}
(\textit{iii}) \textit{comparison between baselines}.
\vspace{-0.3mm}
\begin{itemize} [leftmargin=*]
\item In most cases, BPR is more powerful than BCE \textit{w.r.t.} Top-5 related quality metrics. This suggests that the pairwise ranking BPR enables GCN based CF models to assign user-interested items with high rankings. 
\item The $min$ results mainly occur in BPR and BCE (objective functions based on individual item comparison ), indicating the importance of considering the correlations among items. 
\vspace{-2mm}
\end{itemize}

\begin{table*}[tp]
\centering
  \fontsize{7.4}{9.7}\selectfont
  \caption{Performance comparison between L$k$P and state-of-the-art ranking models, deployed on matrix factorization.}
  \vspace{-0.6mm}
  \setlength{\tabcolsep}{1.5mm}{
  \label{tab:performance_comparison}
    \begin{tabular}{llllllllllllll}
    \Xhline{0.8pt}
    Dataset&Method&Re@5&Re@10&Re@20&Nd@5&Nd@10&Nd@20&CC@5&CC@10&CC@20&F@5&F@10&F@20  \cr\Xhline{0.6pt}
    \multirow{7}{*}[1pt]{Beauty} 
    &L$k$P$_{PS}$-MF &0.0785&0.1265&0.1787&0.0765&\textbf{0.0976}&0.1176&\textbf{0.0569}&\textbf{0.0894}&\textbf{0.1346}&0.0656& \textbf{0.0995}& 0.1409 \\ 
    &L$k$P$_{NPS}$-MF &\textbf{0.0817}&\textbf{0.1285}&\textbf{0.1849}&\textbf{0.0768}&0.0973&\textbf{0.1190}&0.0567&0.0886&0.1438&\textbf{0.0661}& 0.0993& \textbf{0.1422} \\  \cline{2-14}
    &BPR-MF &0.0690\boldmath{$^{-}$}&0.1026\boldmath{$^{-}$}&0.1584\boldmath{$^{-}$}&0.0701\boldmath{$^{-}$}&0.0697\boldmath{$^{-}$}&0.1012\boldmath{$^{-}$}&0.0574\boldmath{$^{+}$}&0.0901\boldmath{$^{+}$}&0.1304\boldmath{$^{+}$}&0.0629\boldmath{$^{-}$}& 0.0881\boldmath{$^{-}$}& 0.1301\boldmath{$^{-}$} \\
    &SetRank-MF &0.0761\boldmath{$^{+}$}&0.1197&0.1689&0.0753\boldmath{$^{+}$}&0.0913&0.1106&0.0567&0.0876&0.1292&0.0648\boldmath{$^{+}$}& 0.0957\boldmath{$^{+}$}& 0.1343\boldmath{$^{+}$} \\
    &S2SRank-MF &0.0757&0.1205\boldmath{$^{+}$}&0.1715\boldmath{$^{+}$}&0.0750&0.0920\boldmath{$^{+}$}&0.1115\boldmath{$^{+}$}&0.0559\boldmath{$^{-}$}&0.0850\boldmath{$^{-}$}&0.1254\boldmath{$^{-}$}&0.0642 & 0.0944& 0.1330 \\
    \cline{2-14}
    &\textit{max vs. min}&18.41&25.24&16.73&9.56&40.03&17.59&1.79&5.18&7.26&5.11& 12.91.& 9.68 \\
    &\textit{max vs. max}&7.36&6.64&7.81&1.99&6.09&6.73&-0.87&-0.78&3.14&1.96& 3.90.& 6.28 \\
    \Xhline{0.6pt}
    \multirow{7}{*}[1pt]{ML}
    &L$k$P$_{PS}$-MF &\textbf{0.0825}&0.1236&\textbf{0.1942}&0.0894&0.1058&0.1239&\textbf{0.3210}&0.5174&0.6526&0.1356 & 0.1878 & 0.2558  \\
    &L$k$P$_{NPS}$-MF &0.0820 &\textbf{0.1286} &0.1933 &\textbf{0.0924} &\textbf{0.1146} &\textbf{0.1265} &0.3207 &\textbf{0.5192} &\textbf{0.6573} &\textbf{0.1371} & \textbf{0.1970} &\textbf{0.2572}\\ \cline{2-14}
    &BPR-MF &0.0726\boldmath{$^{-}$} &0.1107\boldmath{$^{-}$} &0.1850\boldmath{$^{-}$} &0.0814\boldmath{$^{-}$} &0.0943\boldmath{$^{-}$} &0.1167\boldmath{$^{-}$} &0.3195\boldmath{$^{+}$} & 0.5180\boldmath{$^{+}$} &0.6505\boldmath{$^{+}$} &0.1239\boldmath{$^{-}$} & 0.1710\boldmath{$^{-}$} & 0.2449\boldmath{$^{-}$} \\
    &SetRank-MF &0.0780\boldmath{$^{+}$} &0.1169 &0.1903\boldmath{$^{+}$} &0.0865 &0.1029 &0.1176 &0.3161\boldmath{$^{-}$} &0.5126\boldmath{$^{-}$} &0.6483 &0.1302\boldmath{$^{+}$} & 0.1810 & 0.2488  \\ 
    &S2SRank-MF &0.0761 &0.1231\boldmath{$^{+}$} &0.1897 &0.0870\boldmath{$^{+}$} &0.1052\boldmath{$^{+}$} &0.1234\boldmath{$^{+}$} &0.3168 &0.5139 &0.6480\boldmath{$^{-}$} &0.1297 & 0.1868\boldmath{$^{+}$} & 0.2522\boldmath{$^{+}$} \\
    \cline{2-14}
    &\textit{max vs. min}&13.64 & 16.17 & 4.97 & 13.51 & 21.53 & 8.40 & 2.85 & 1.29 & 1.44 & 10.70 & 15.22 & 5.03 \\
    &\textit{max vs. max}&5.77 & 4.47 & 2.05 & 6.21 & 8.94 & 2.51 & 1.33 & 0.62 & 1.05 & 5.32 & 5.48 & 2.00 \\
    \Xhline{0.6pt}
    \multirow{7}{*}[1pt]{Anime} 
    &L$k$P$_{PS}$-MF &\textbf{0.0887}&0.1490&0.2427&\textbf{0.1450}&\textbf{0.1509}&0.1942&0.3012&0.4236&0.5504&\textbf{0.1684}& \textbf{0.2215}& 0.3128 \\
    &L$k$P$_{NPS}$-MF &0.0875&\textbf{0.1497}&\textbf{0.2438}&0.1425&0.1493&\textbf{0.1950}&\textbf{0.3090}&\textbf{0.4260}&\textbf{0.5553}&0.1676& 0.2213& \textbf{0.3145} \\ \cline{2-14}
    &BPR-MF &0.0791\boldmath{$^{-}$}&0.1349\boldmath{$^{-}$}&0.2157\boldmath{$^{-}$}&0.1306\boldmath{$^{-}$}&0.1342\boldmath{$^{-}$}&0.1720\boldmath{$^{-}$}&0.3143\boldmath{$^{+}$}&0.4215&0.5453\boldmath{$^{-}$}&0.1554\boldmath{$^{-}$}& 0.2040\boldmath{$^{-}$}& 0.2860\boldmath{$^{-}$} \\
    &SetRank-MF &0.0833\boldmath{$^{+}$}&0.1403&0.2245&0.1358\boldmath{$^{+}$}&0.1396\boldmath{$^{+}$}&0.1848&0.3101&0.4234\boldmath{$^{+}$}&0.5501\boldmath{$^{+}$}&0.1619\boldmath{$^{+}$}& 0.2104\boldmath{$^{+}$}& 0.2983 \\
    &S2SRank-MF &0.0827&0.1420\boldmath{$^{+}$}&0.2268\boldmath{$^{+}$}&0.1349&0.1365&0.1850\boldmath{$^{+}$}&0.3073\boldmath{$^{-}$}&0.4199\boldmath{$^{-}$}&0.5461&0.1608& 0.2091& 0.2990\boldmath{$^{+}$} \\
    \cline{2-14}
    &\textit{max vs. min}&12.14&10.97&13.03&11.03&12.44&13.37&0.57&1.45&1.83&8.33& 8.58& 9.97 \\
    &\textit{max vs. max}&6.48&5.42&7.50&6.77&8.09&5.41&-1.66&0.61&0.95&4.00& 5.29& 5.18 \\
    \Xhline{0.8pt}
    \end{tabular}}
    \label{tab-k-dpp:MF}
    \vspace{-0.2mm}
\end{table*}

To further demonstrate the effectiveness of L$k$P approaches, we conduct more experiments on the basic MF implementation (\textit{i.e.}, directly optimizing the optimization criterion).
According to the comparison in \Cref{tab-k-dpp:GCN}, only the two main and most noteworthy variants (PS and NP) of L$k$P are further compared, denoted by L$k$P$_{PS}$ and L$k$P$_{NPS}$.
Following previous CF studies \cite{wang2020setrank, chen2021set2setrank, wan2022cross}, only the ranking optimization approaches (\textit{i.e.}, BPR, SetRank, S2SRank) are compared in the implementation of MF. 
The corresponding comparison is presented in \Cref{tab-k-dpp:MF}. All experiments are conducted directly following basic MF implementation.
The bold values and superscripts denote the same meanings as those in \Cref{tab-k-dpp:GCN}.

The main findings are: (\romannumeral1) NPS type of L$k$P is generally more powerful than PS type \textit{w.r.t.} quality and diversity evaluation metrics, which reinforces the meaning of taking the items that are of less interest to users in ranking comparisons. 
%%A trend similar to the comparison of \Cref{tab-k-dpp:GCN} can be found here. That is, L$k$P$_{PS}$ beats L$k$P$_{NPS}$ when evaluating the relevance of the higher ranking items (\textit{e.g.}, Top-5);
%(\romannumeral2) On the diversity side, L$k$P$_{PS}$ is generally better than L$k$P$_{NPS}$ on Beauty. The reason behind these results might be that only considering the diversity correlations of target subsets makes L$k$P focus more on handling the complex category distributions, as Beauty products contain more categories than Anime and ML have;
(\romannumeral2) In some cases, BPR achieves the best diversity performance but the worst relevance results among the compared methods, \textit{e.g.}, CC@5 on Beauty and Anime. This might be caused by the application of basic MF, as the simple MF endows embeddings learned from BPR with more randomness compared to other set-level ranking optimization approaches, and therefore diversity is promoted but relevance is sacrificed;
(\romannumeral3) Compared to SetRank and S2SRank that consider the correlations of item sets, around 5\% improvement is achieved on overall performance (F). This again verifies the effectiveness of our approaches even deploying them on basic and simple MF. Aligning with analysis from GCN and MF implementations, additional and extensive evidence can be provided to support the superiority of our approaches.

%%The main observations from the MF implementations are similar to those of Table 1. Detailed results and analysis are presented in Table 4 of Appendix A.5. 

%%without normalization, the results are not good and easily have mistake. show the importance of ranking implication.

%%improvement of kdpp over original model, demonstrating our ranking method can be easily deployed on existing models and achieve significant improvements
%%%% lightgcn beats gcn, our designed gcn based model outperforms lightgcn means the proposed loss endows powerful ability for recommendation. 
%%% use superscript * for demonstration 
\begin{table*}[tp]
\centering
  \fontsize{7.4}{9.7}\selectfont
  \caption{Performance comparison between strong baselines and their $k$-DPP reworked counterparts.}
  \vspace{-0.6mm}
  \setlength{\tabcolsep}{2.2mm}{
  \label{tab:performance_comparison}
    \begin{tabular}{llcccccccccccc}
    \Xhline{0.8pt}
    Dataset&Method&Re@5&Re@10&Re@20&Nd@5&Nd@10&Nd@20&CC@5&CC@10&CC@20&F@5&F@10&F@20  \cr\Xhline{0.6pt}
    \multirow{8}{*}[1pt]{Beauty} 
    &GCMC &0.0767&0.1077&0.1609&0.0790&0.1054&0.1215&0.0590&0.0901&0.1320&0.0671& 0.0976& 0.1364 \\ 
    &GCMC$_{PS}$ &\textbf{0.0794}&0.1106&0.1655&0.0813&0.1120&0.1307&0.0582&0.0922&0.1356&0.0675& 0.1009& 0.1416 \\
    &GCMC$_{NPS}$ &0.0792&\textbf{0.1174}&\textbf{0.1706}&\textbf{0.0816}&\textbf{0.1135}&\textbf{0.1334}&\textbf{0.0592}&\textbf{0.0926}&\textbf{0.1371}&\textbf{0.0685}& \textbf{0.1028}& \textbf{0.1442} \\ \cline{2-14}
    &\textit{Improv} (\%) &3.52& 9.01&6.03 &3.29 &7.69 &9.79 &1.02 &2.77 &3.86 &1.98 & 5.26
 &5.66  \\ \cline{2-14}
    &NeuMF &0.0781&0.1254&0.1689&0.0804&0.1020&0.1163&0.0557&0.0861&0.1273&0.0654& 0.0980& 0.1345 \\
    &NeuMF$_{PS}$ &0.0790&0.1250&0.1724&\textbf{0.0823}&0.1025&0.1185&\textbf{0.0562}&0.0870&0.1285&\textbf{0.0662}& 0.0986& 0.1365 \\
    &NeuMF$_{NPS}$ &\textbf{0.0793}&\textbf{0.1259}&\textbf{0.1750}&0.0811&\textbf{0.1033}&\textbf{0.1190}&0.0560&\textbf{0.0875}&\textbf{0.1286}&0.0660& \textbf{0.0993}& \textbf{0.1372} \\ \cline{2-14}
     &\textit{Improv} (\%) &1.54 &0.40 &3.61 &2.36 &1.27 &2.32 &0.90 &1.74  &1.02 &1.25 &1.33 &1.98 \\
    \Xhline{0.6pt}
    \multirow{8}{*}[1pt]{ML}
    &GCMC &0.0800 &0.1176 &0.1826 &0.0920 &0.1237 &0.1494 &0.3506 &0.5317 &0.6726 &0.1381 & 0.1967 & 0.2663  \\
    &GCMC$_{PS}$ &0.0856 &0.1231 &0.1842 &0.0964 &0.1251 &0.1527 &0.3546 &0.5320 &0.6740 &0.1448 & 0.2013 & 0.2695 \\ 
    &GCMC$_{NPS}$ &\textbf{0.0861} &\textbf{0.1235} &\textbf{0.1870} &\textbf{0.0982} &\textbf{0.1279} &\textbf{0.1552} &\textbf{0.3552} &\textbf{0.5374} &\textbf{0.6772} &\textbf{0.1463} & \textbf{0.2037} & \textbf{0.2732}\\  \cline{2-14}
     &\textit{Improv} (\%) &7.63. & 5.02 & 2.41 & 6.74 & 3.40 & 3.88 & 1.31 & 1.07 & 0.68 & 5.94 & 3.60 & 2.59  \\
    \cline{2-14}
    &NeuMF &0.0760 &0.1092 &0.1747 &0.0867 &0.1109 &0.1279 &0.3458 &0.5397 &0.6684 &0.1317 & 0.1828 & 0.2467 \\
    &NeuMF$_{PS}$ &\textbf{0.0806} &0.1157 &0.1764 &0.0884 &0.1138 &0.1293 &\textbf{0.3502} &0.5472 &0.6704 &0.1361 & 0.1897 & 0.2489 \\
    &NeuMF$_{NPS}$ &0.0792 &\textbf{0.1160} &\textbf{0.1775} &\textbf{0.0902} &\textbf{0.1145} &\textbf{0.1311} &0.3479 &\textbf{0.5484} &\textbf{0.6715} &\textbf{0.1362} & \textbf{0.1905} & \textbf{0.2510} \\ \cline{2-14}
     &\textit{Improv} (\%) &6.05 & 6.23 & 1.60 & 4.04 & 3.25 & 2.50 & 1.27 & 1.61 & 0.46 & 3.43 & 4.18 & 1.70  \\
     \Xhline{0.6pt}
    \multirow{8}{*}[1pt]{Anime} 
    &GCMC &0.0883&0.1387&0.2199&0.1521&0.1743&0.2150&0.3508&\textbf{0.4625}&0.5749&0.1790& 0.2339& 0.3155 \\
    &GCMC$_{PS}$ &0.0910&\textbf{0.1473}&0.2315&\textbf{0.1579}&0.1755&0.2216&0.3496&0.4601&0.5760&\textbf{0.1836}& 0.2390& 0.3252 \\
    &GCMC$_{NPS}$ &\textbf{0.0914}&0.1471&\textbf{0.2326}&0.1567&\textbf{0.1763}&\textbf{0.2232}&\textbf{0.3513}&0.4620&\textbf{0.5765}&0.1834& \textbf{0.2396}& \textbf{0.3267} \\ \cline{2-14}
     &\textit{Improv} (\%) &3.51 &6.20 &5.78 &3.81 &1.15 &3.81 &0.14 &-0.11 &0.28 &2.52 &2.43 
 &3.52  \\ \cline{2-14}
    &NeuMF &0.0710&0.1229&0.1985&0.1142&0.1219&0.1576&0.3829&0.4919&0.6099&0.1491& 0.1960& 0.2756 \\
    &NeuMF$_{PS}$ &0.0750&0.1239&0.2027&0.1190&0.1239&0.1611&0.3836&0.4930&0.6106&0.1548& 0.1980& 0.2803 \\
    &NeuMF$_{NPS}$ &\textbf{0.0757}&\textbf{0.1252}&\textbf{0.2042}&\textbf{0.1198}&\textbf{0.1250}&\textbf{0.1621}&\textbf{0.3838}&\textbf{0.4944}&\textbf{0.6107}&\textbf{0.1559}& \textbf{0.1997}& \textbf{0.2818} \\ \cline{2-14}
     &\textit{Improv} (\%) &6.62 &1.87 &2.87 &4.90 &2.54 &2.86 &0.39 &0.51 &0.13 &4.51 &1.86 &2.23 \\
     
    \Xhline{0.8pt}
    \end{tabular}}
    \label{tab-k-dpp:seminal}
    \vspace{-3mm}
\end{table*}

To evaluate the adaptability and generality of $k$-DPP based approaches for the item recommendation area, we deploy two main types of L$k$P (PS and NPS) on popular CF models (\textit{i.e.}, GCMC and NeuMF) by replacing their original recommendation objective function. If the reworked models that apply L$k$P obtain better performance than the original frameworks, we can validate two advantages of this work: (\romannumeral1) L$k$P can be adaptively applied to existing CF models as an objective function, \textit{i.e.}, can be generalized to distinct recommendation models; (\romannumeral2) the boosted performance is another evidence to show the superiority of L$k$P.

The corresponding comparison is shown in \Cref{tab-k-dpp:seminal}, where the methods with subscripts $_{PS}$ and $_{NPS}$ denote the reworked models by L$k$P$_{PS}$ and L$k$P$_{NPS}$, respectively. 
The bold value is the best among three methods (a baseline and two reworked models on a dataset), since we only consider the improvement achieved by reworked models over the original baseline, the comparison between different baselines is not needed. 
\textit{Improv}(\%) shows how much L$k$P (the better approach out of PS and NPS) improves over the original baseline. We can find that (\textit{i}) In most cases, obvious improvements (particularly on beauty dataset) are obtained by the reworked models compared to original baselines on both quality and diversity evaluation metrics; 
(\textit{ii}) In the comparison between L$k$P$_{PS}$ and L$k$P$_{NPS}$, NPS type provides better suggestions (more relevant and diversified) for users. This finding is consistent with the results of \Cref{tab-k-dpp:GCN} and \Cref{tab-k-dpp:MF}.

Overall, the reworked models (\Cref{tab-k-dpp:seminal}) and GCN-based implementations (\Cref{tab-k-dpp:GCN}) exhibit more pronounced improvements than the basic MF-based methods. This could be attributed to the fact that the simplistic MF might not possess the same representation learning capacity as the neural network-based methods, thereby hindering their ability to fully exploit the relationships captured by set-level ranking optimization. Nevertheless, our approaches consistently demonstrates over a 5\% enhancement on both basic and recently proposed neural models in most scenarios. This highlights that L$k$P, which perceives multiple items as a cohesive entity and employs probability distributions for relevance and diversity ranking, is not only effective and but also propels advancements of both recent and classic techniques within the domain.

\subsubsection{Analysis}

\begin{figure}
\centering
\captionsetup[subfloat]{farskip=0pt, captionskip=-1pt} 
    \subfloat[L$k$P$_{PS}$]{
    \includegraphics[width=0.45\linewidth]{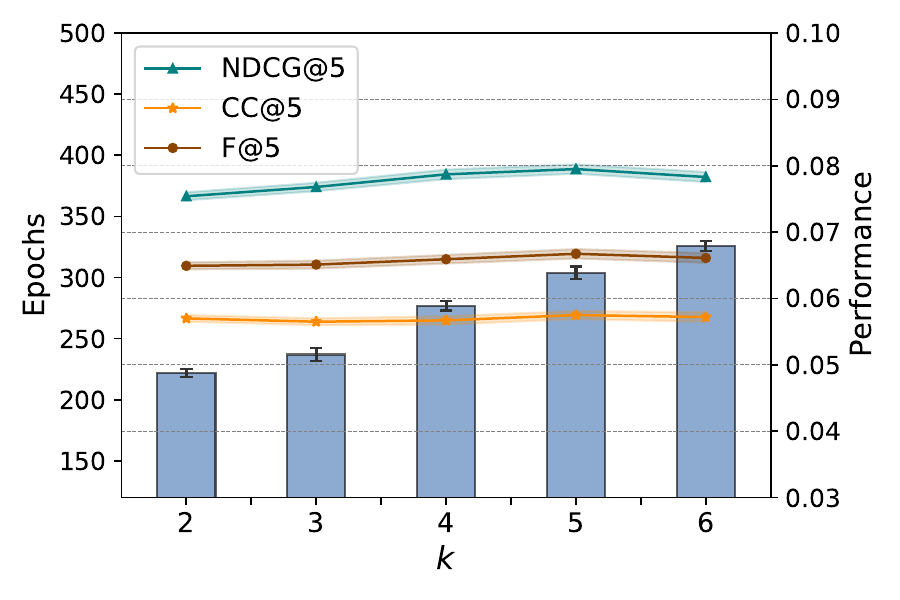}
    \label{fig-k-dpp:ps-nps-k:sub1}
    }
    \hfill
    \subfloat[L$k$P$_{NPS}$]{
    \includegraphics[width=0.45\linewidth]{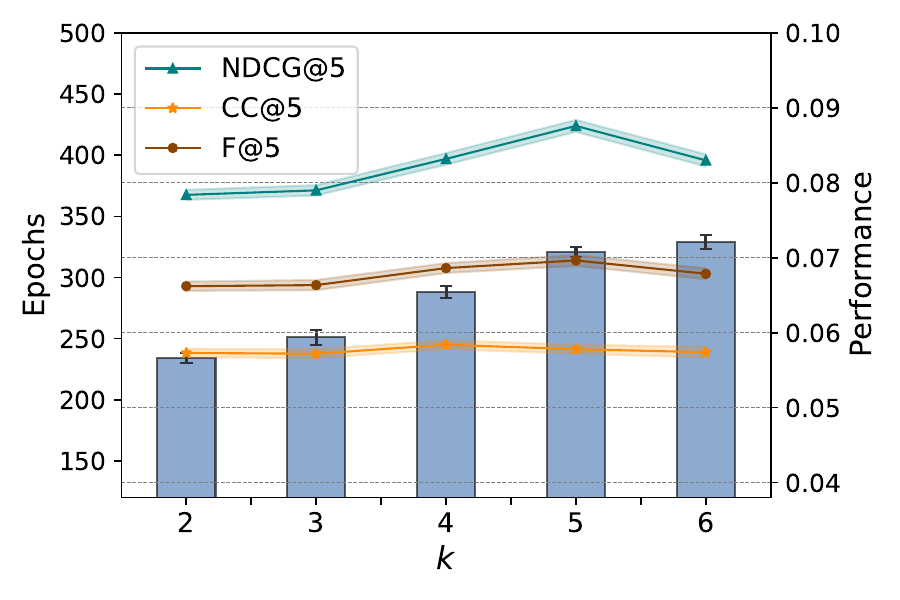}
    \label{fig-k-dpp:ps-nps-k:sub2}
    }
%%    \\
%%    \subfloat[\textcolor{blue}{L$k$P$_{PS}$ on Anime}]{
%%    \includegraphics[width=0.44\linewidth]{ICDE-kdpp-ps-k-anime.pdf}
%%    \label{fig-k-dpp:ps-nps-k:sub3}
%%   }
%%    \hfill
%%    \subfloat[\textcolor{blue}{L$k$P$_{NPS}$ on Anime}]{
%%    \includegraphics[width=0.44\linewidth]{ICDE-kdpp-nps-k-anime.pdf}
%%    \label{fig-k-dpp:ps-nps-k:sub4}%%}
\vspace{-1mm}
\caption{Performance trends at different $k$ on Beauty.}
\label{fig-k-dpp:ps-nps-k}
\vspace{-2.5mm}
\end{figure}

\begin{figure}
\centering
\captionsetup[subfloat]{farskip=0pt, captionskip=-1pt} 
    \subfloat[Top-5 performance]{
    \includegraphics[width=0.45\linewidth]{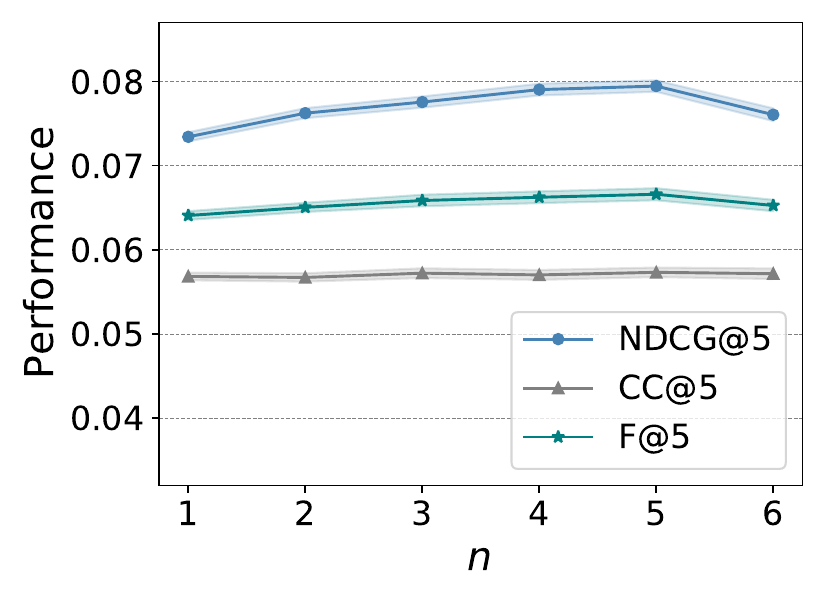}
    \label{fig-k-dpp:ps-n:sub1}
    }
    \hfill
    \subfloat[Top-20 performance]{
    \includegraphics[width=0.45\linewidth]{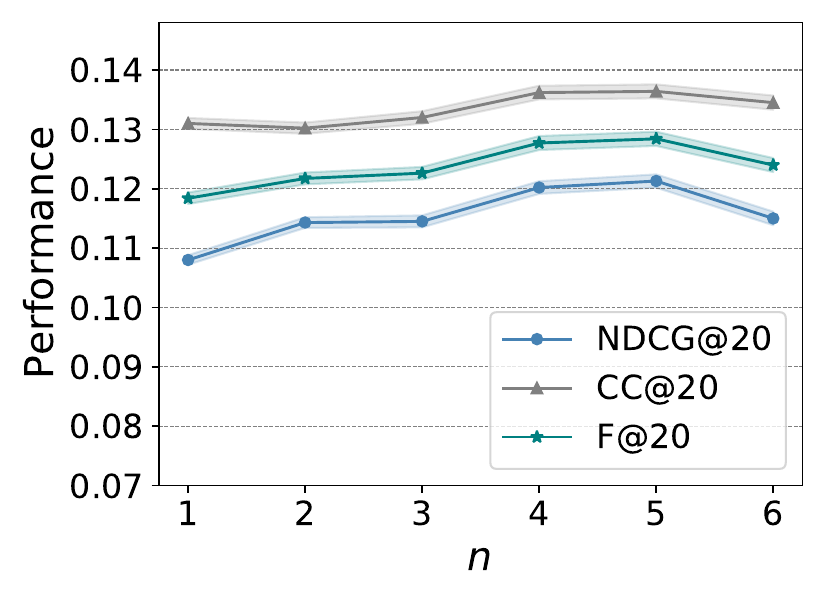}
    \label{fig-k-dpp:ps-n:sub2}
    }
\vspace{-0.6mm}
\caption{Performance of L$k$P$_{PS}$ at different $n$.}
\label{fig-k-dpp:ps-n}
\vspace{-3.5mm}
\end{figure}

\Cref{tab-k-dpp:GCN}-\Cref{tab-k-dpp:seminal} have verified the superiority of L$k$P from multiple aspects. We now further discuss the attributes of our approaches.
%%by analyzing the impact of some important parameters. 
%%As mentioned in \Cref{sec-k-dpp:k-dpp-probability}, the tailored ground set setting plays a vital role in maintaining the \textit{ranking interpretation}. 
Experiments in \Cref{tab-k-dpp:GCN}-\Cref{tab-k-dpp:seminal} are conducted under the same setting of $k=n=5$. 
%%We therefore perform additional experiments by fixing all parameters except $k$ and $n$. 
We use \Cref{fig-k-dpp:ps-nps-k} and \Cref{fig-k-dpp:ps-n} to illustrate the effects of $k$ and $n$
%%, \textit{i.e.}, number of observed items in the specific ground set, 
for CF recommendations. 
Note that the error bars in the figures represent the standard errors. The slightly visible error bars confirm the stability of our experiments.
%%Figure (a) and (b) present the trends of L$k$P$_{PS}$ and L$k$P$_{NPS}$ on Beauty dataset. $n$ is set to equal $k$ in the corresponding experiments. L$k$P$_{PS}$ and L$k$P$_{NPS}$ results on Anime are shown in Figure (c) and (d). 
Besides the performance of three evaluation metrics, the epochs that are needed by L$k$P$_{PS}$ or L$k$P$_{NPS}$ to achieve the best performance under each setting are also presented. 
%The dotted line on sub-figures is used to indicate the best trade-off performance for comparison.  
As L$k$P regards multiple items as a whole and tries to capture correlations among them, we therefore only perform experiments on $k>1$.  
The slight decrease of CC results when $k>4$ might be caused by the correlations among items, as $k$-DPP enjoys more correlations for recommendation relevance but lowers the weight of diversity.  
The quality performance (NDCG@5) clearly increases at first when $k<5$, but a downward trend happens when $k>5$. This suggests that if excessive correlations among items are considered, the final performance will be negatively affected. 
As for the epochs, an obvious trend, \textit{i.e.}, an increase with $k$, can be found, which is straight-forward to explain as $k$-DPP needing more time to learn the more complex distribution. 

\Cref{fig-k-dpp:ps-n} shows L$k$P$_{PS}$ performance (Top-5 and Top-20) on Beauty with different numbers of unobserved items (\textit{i.e.}, $n$). We find that all three metrics at different rankings show a similar trend. That is, they smoothly increase at first and then decrease. These results mean that the moderate size of $n$ provides a sufficient set-level ranking comparison for learning $k$-DPP probabilities. 
However, when redundant comparisons are provided (\textit{e.g.}, $n>5$), it becomes difficult for $k$-DPP to capture effective correlations. 
%%and to calculate subset probabilities. 
As mentioned in \Cref{sec-k-dpp:k-dpp-probability}, we set $k=n$ and the number of targets in the ground set equals to $k$ for L$k$P$_{NPS}$ and L$k$P$_{PS}$ to keep the ranking interpretation. 
External experiments in the cases of $n>k$ on NPS and the number of targets (6 and 7) $> k$ (5) on PS have been performed. However, the worse performance compared to the presented cases suggests the importance of maintaining the ranking interpretation. Experiments in \Cref{fig-k-dpp:ps-nps-k} and \Cref{fig-k-dpp:ps-n} are conducted on GCN implementation. Similar conclusions are obtained from experiments deployed on MF. 

%if k!=n???
%why ILD is better as it is not directly related to category???
\begin{figure}
\centering
\captionsetup[subfloat]{farskip=0pt, captionskip=-1pt} 
    \subfloat[L$k$P$_{PS}$]{
    \includegraphics[width=0.44\linewidth]{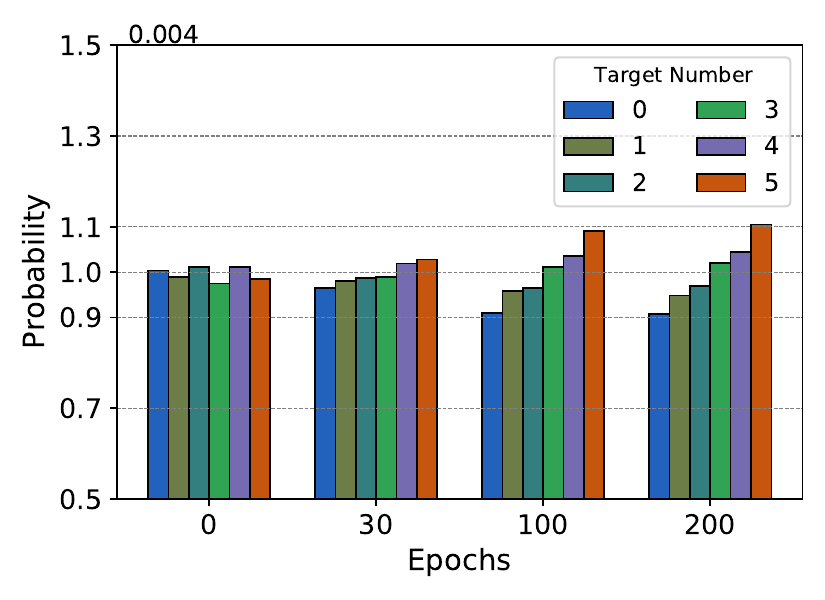}
    \label{fig-k-dpp:examples:sub1}
    }
    \hfill
    \subfloat[L$k$P$_{NPS}$]{
    \includegraphics[width=0.44\linewidth]{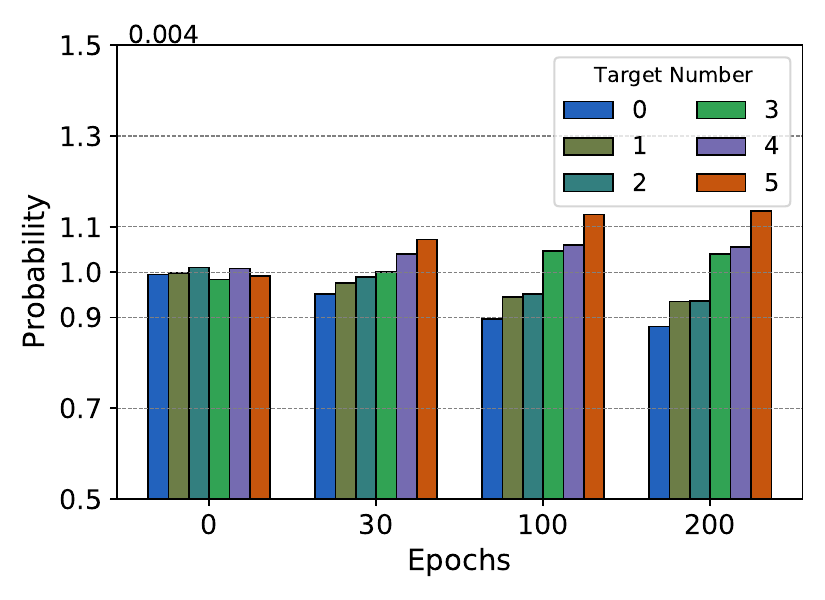}
    \label{fig-k-dpp:examples:sub2}
    }
\caption{Probability distributions at different epochs on Anime.}
\vspace{-4mm}
\label{fig-k-dpp:examples}
\end{figure}   

%%\begin{figure}
%%\centering
%%    \subfigure[L$k$P$_{PS}$]{
%%    \includegraphics[width=0.47\linewidth]{LkP-ps-sets-probability.pdf}}
%%    \subfigure[L$k$P$_{NPS}$]{
%%    \includegraphics[width=0.47\linewidth]{LkP-nps-sets-probability.pdf}}
%%    \vspace{-3mm}
%%\caption{Subset ranking of L$k$P$_{PS}$ and L$k$P$_{NPS}$ at different epochs on Anime.}
%%\label{fig-k-dpp:examples}
%%\vspace{-6mm}
%%\end{figure}  

%%\begin{figure*}
%%  \centering
%%  \includegraphics[width=0.93\linewidth]{k-DPP_ranking-example.png}
%%  \vspace{-4mm}
%%  \caption{Examples of subset probability ranking of L$k$P$_{PS}$ and L$k$P$_{NPS}$ at different epochs on Anime.}
%%  \label{fig-k-dpp:examples}
%%  \vspace{-4mm}
%% \end{figure*}

To diagrammatically present the ranking interpretation, we use real examples to show the probability ranking of $k$-DPP subsets.
We randomly select 100 training instances ($k+n$ ground set where $k=n=5$) in the learning process, and divide all $k$-sized subsets (252) of a ground set into six groups according to the target number in a subset, \textit{e.g.}, the group of $k$-subsets with 1 target comprised of 25 subsets (the subsets filled by 1 observed item and 4 unobserved ones), and target number $=5$ represents the target subset $S_u^{+k}$ drawn from a $k$-DPP formulated in \Cref{equ-k-dpp:L-k+n-positive}. 
The $k$ targets and $n$ random items of training instances contain the same number of categories.
%%To clearly illustrate the trend of probability ranking, we average the $k$-DPP probabilities of subsets in the same group. 
%%The main observation made based on the results is that with the increase of target number in subsets, the target $k$-DPP probability ascends gradually as training proceeds, meaning the ranking implications are demonstrated, \textit{i.e.}, the target subset ranks higher than other subsets with cardinality $k$. 

\Cref{fig-k-dpp:examples} presents probability ranking trends of L$k$P$_{PS}$ and L$k$P$_{NPS}$ employing GCN at 0, 30, 100, and 200 training epochs on Beauty.
To illustrate the ranking interpretation of $k$-DPP probability learning, probabilities of 252 $k$-sized subsets from each $5+5$ ground set, \textit{i.e.}, $C(5+5, 5)$ combinations, are calculated. 
Before the $k$-DPP learning process starts, randomly distributed subsets (cardinality $k$) are assumed to have the equivalent probability of being selected as a $k$-DPP, \textit{i.e.}, 0.004 (1/252) for each subset.
Each bar in a plot denotes the averaged probabilities of subsets that contain the same number of observed items from randomly selected training instances.

We can see that subsets with differing numbers of targets indeed have close normalized $k$-DPP probabilities (\textit{i.e.}, near 0.004) when epoch $=0$. 
With the learning process proceeding, the discrepancy between $k$-DPP probabilities of subsets with different target numbers obviously grows at first, \textit{i.e.}, subsets with more targets are endowed with higher probabilities, and the trend tends to be stable when epochs reach 200. 
These illustrate that 
%L$k$P approaches learn to rank subsets (with cardinality $k$) according to the probability of multiple items being selected as satisfied recommendations. That is, 
\textit{relevance ranking interpretation} behind L$k$P are validated by the $k$-DPP probability examples, as the target subset ranks higher (with greater $k$-DPP probability) than other subsets, and the subset with unobserved items ranks lower than the subsets that contain one or more targets.
%we only display the examples on Beauty using GCN.
In addition, considering the exclusion of negative subsets in $k$-DPP (\textit{i.e.}, L$k$P$_{NPS}$) facilitates widening the ranking gap between target set and unobserved item set compared to L$k$P$_{NS}$, further enhancing performance improvements (as shown in \Cref{tab-k-dpp:GCN}-\Cref{tab-k-dpp:seminal}). The similar trends can be found on other datasets and MF implementation. 

We also conduct a comparison between the averaged $k$-DPP probabilities ($k$=$n$=5) of two types of target subsets on Anime using L$k$P$_{PS}$, \textit{i.e.}, monotonous target subsets with limited categories ($<4$ categories) and diversified target subsets with broader categories ($>5$ categories). The observed average probabilities of 0.0041 \textit{vs.} 0.0040, 0.0045 \textit{vs.} 0.0042, and 0.0046 \textit{vs.} 0.0043 for diversified and monotonous target subsets at 0, 100, and 200 epochs respectively, confirm the significant role of \textit{diversity ranking interpretation} across $(k+n)$ $k$-DPP distributions in our optimization approaches. Because the diverse kernel $\mathbf{K}$ is pre-learned, on average, diverse target sets from half of randomly selected training instances ($k$-DPP distributions) have a relatively higher probability than other monotonous target sets from the $k$-DPP distributions of the remaining half, even before the L$k$P optimization begins.

These results again indicate the importance of $k$-DPP normalization, as only in this case the ranking interpretation is provided. We also perform experiments in the case of removing normalization of $k$-DPP, and then two serious problems occur: (\romannumeral1) the deep learning frameworks (both Pytorch and TensorFlow) cannot handle the complex gradient calculation of the DPP, as the non-normalized determinant values are unstable; 
(\romannumeral2) even if we use some techniques, \textit{e.g.}, normalizing diverse kernel or adding activation function to control the values of relevance scores, to avoid the gradient calculation error, the final recommendation results still cannot compare with L$k$P approaches, as the important ranking interpretation for personalized ranking is ignored. 
These also elucidate why the standard DPP is inapplicable for ranking-based optimization criterion. Within the standard DPP distribution, ranking competition arises among subsets of all sizes, leading to an uninterpretable and ineffective ranking optimization (concrete analysis in \Cref{sec-k-dpp:relevance-comparison}). According to our experiments under the same setting as \Cref{tab-k-dpp:GCN}-\Cref{tab-k-dpp:seminal}, applying standard DPP for the ranking probability formulation (\Cref{equ-k-dpp:L-k+n-positive}) achieves an unacceptable performance that is weaker than BPR (0.1106 \textit{vs.} 0.1254 at NDCG@20 on ML). This is the reason we omit the results based on standard DPP in this section.
%%% Further explanation has been detailed in the third part of \Cref{sec-k-dpp:related-work}.

\begin{figure}
  \centering
  \includegraphics[width=0.95\linewidth]{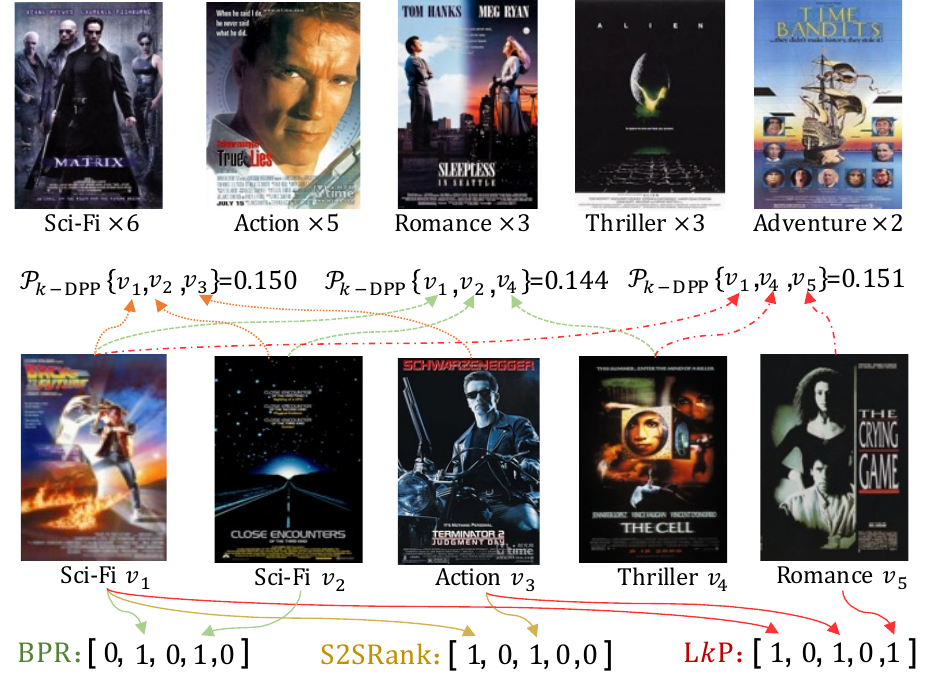}
  \vspace{-2mm}
  \caption{Case study of L$k$P$_{PS}$ optimization criterion.}
  \label{fig-k-dpp:case-study}
  \vspace{-4mm}
\end{figure}

In \Cref{fig-k-dpp:case-study}, we conduct a case study on $u_{1518}$ of ML-1M using GCN-based L$k$P$_{PS}$. For clarity, we categorize each movie reviewed by $u_{1518}$ into a key genre based on its synopsis on Wikipedia. Five genres emerge in the training data, while the test data comprises five movies across four genres. Training movies reveal the user's inclination towards Sci-Fi (6) and Action (5) movies. Both BPR and S2SRank methods acknowledge this preference by recommending target movies (denoted as 1) in the predicted Top 5 list at positions 2 and 4, and 1 and 3 respectively. Our L$k$P method, while recognizing the user's apparent preferences, also considers diversity by suggesting a hidden target romance movie. This highlights L$k$P's ability to cater to relevance and diversity. We also analyze $k$-DPP probabilities for 3-sized subsets over movies for testing and observe: (\textit{\romannumeral1}) The diversified item subset $\{v_1, v_4, v_5\}$ has a higher probability than monotonous ones, affirming our consideration on diversity comparison; (\textit{\romannumeral2}) Despite similar category scopes, the subset $\{v_1, v_2, v_3\}$ possesses a higher probability than $\{v_1, v_2, v_4\}$, which suggests that a stronger dependency between $v_3$ and $\{v_1, v_2\}$ leads to a higher predictive value for $v_3$, further emphasizing the capability of concrete set-level ranking to capture intricate relationships. 

%%\begin{figure}
%%\centering
%%    \subfigure[Beauty]{
%%    \includegraphics[width=0.42\linewidth]{ICDE-kdpp-time-beauty.pdf}}
%%    \subfigure[Anime]{
%%    \includegraphics[width=0.42\linewidth]{ICDE-kdpp-time-anime.pdf}}
%%    \vspace{-3mm}
%%\caption{Training efficiency comparison.}
%5\label{fig-k-dpp:time}
%%\vspace{-5mm}
%%\end{figure}

%test the probability of different sets (examples)
%probability of k-DPP-distributed subset; use a figure to compare all subsets' probabilities, using a scatter diagram , whether Y+ ranks higher than all the others
%model framework constrains the performance
%the importance of normalization. 
%\textbf{Collaborative Filtering.}
%The main meaning of CF is to model users’ preference on items based on historical interactions \cite{he2017neural, wang2019unified, liu2010unifying}. Among the various collaborative filtering techniques, MF and neural CF are most popular. MF based models learn user and item embedddings based on low-rank decomposition of the user-item interaction matrix \cite{koren2008factorization, koren2009matrix}.The huge success of neural networks encourages researchers to represent the inner product with a neural architecture \cite{he2017neural}. Some studies treated the user-item behaviors as a user-item bipartite graph, and designed neural graph models \cite{wang2019neural, he2020lightgcn, chen2020revisiting}, in which the crucial collaborative signals are distill. 
%We list the studies that are closely related to ours.
\vspace{-1mm}
\section{Conclusion}
\vspace{-1mm}
In this work, we propose a new optimization criterion L$k$P based on $k$-DPP set probability comparison for personalized ranking focusing on implicit feedback, enlightening a new perspective (set-level probability comparison) in the formulation of ranking optimization. 
Based on this, two ranking optimization approaches L$k$P$_{PS}$ and L$k$P$_{NPS}$ are designed that take into account different ranking considerations. 
As such, multiple observed items and unobserved items are directly regarded as subsets with fixed cardinality $k$, and are compared through the ranking interpretation based on the tailored $k$-DPP distribution, notably with probabilistic normalization that respects the cardinality constraint. 
Three levels of evaluation and comparison in the experiments comprehensively demonstrate the superiority of our approaches, and the ranking interpretation is validated through real-world examples. Notably, plugging our optimization criterion into strong baselines provides a ready source of improved performance.
Experimental results help us to analyze the different advantages of two L$k$P approaches and provide insight into making selection between them in different circumstances. 
L$k$P$_{PS}$ and L$k$P$_{NPS}$ also have potential in ranking related tasks, such as Web search \cite{leung2010personalized, kasneci2008naga} and spam detection \cite{chirita2005mailrank, saini2019multi, liu2020recommending}, which will be explored in the future.

\bibliography{reference}{}
\bibliographystyle{IEEEtran}

\end{document}